\documentclass[aps,prl,reprint,superscriptaddress,amsmath,amssymb,showpacs,floatfix,longbibliography]{revtex4-1}

\usepackage{graphicx}
\usepackage{dcolumn}
\usepackage{bm}
\usepackage{color}
\usepackage[utf8]{inputenc}
\usepackage{physics}
\usepackage{mathtools}
\usepackage{soul}
\usepackage{xcolor}
\usepackage{booktabs}

\begin{document}

\title{
Fluctuation-Response Theory for Nonequilibrium Langevin Dynamics}

\author{Hyun-Myung Chun}
\affiliation{School of Physics, Korea Institute for Advanced Study, Seoul 02455, Republic of Korea}
\author{Euijoon Kwon}
\affiliation{School of Physics, Korea Institute for Advanced Study, Seoul 02455, Republic of Korea}
\affiliation{Department of Physics and Astronomy \& Center for Theoretical Physics, Seoul National University, Seoul 08826, Republic of Korea}
\author{Hyunggyu Park}
\affiliation{Quantum Universe Center, Korea Institute for Advanced Study, Seoul 02455, Korea}
\author{Jae Sung Lee}
\email{jslee@kias.re.kr}
\affiliation{School of Physics, Korea Institute for Advanced Study, Seoul 02455, Republic of Korea}

\date{\today}

\begin{abstract}
{We establish a unified fluctuation–response relation for Langevin dynamics. By exploiting the common mathematical structures underlying fluctuations and responses of empirical density and current, we derive a unified identity that generalizes the fluctuation–dissipation theorem from equilibrium to nonequilibrium settings. This relation connects global fluctuations of observables with their local responses to perturbations in force, mobility, and temperature. We further derive finite-time fluctuation–response inequalities, leading to response uncertainty relations that complement the identity by providing more practical bounds. These derivations establish a unified theoretical framework linking the fluctuation–dissipation theorem and thermodynamic uncertainty relations. Using the $F_1$-ATPase molecular motor model, we illustrate how these response-based bounds constrain the long-time diffusion coefficient.}
\end{abstract}

\maketitle

%%%%%%%%%%%%%%%%%%%%%%%%%%%%%%%%%%%%%%%%%%%%%%%%%%%%%%%%%%%%%%%%%%%%%%%%%%%%%
\emph{Introduction}-- Understanding the relationship between fluctuations and response is a central theme in statistical physics. Near equilibrium, this relationship is firmly established by the seminal fluctuation-dissipation theorem (FDT), which expresses the linear response of a system in terms of its equilibrium fluctuations~\cite{kubo1966fluctuation,kubo2012statistical,marconi2008fluctuation}. However, far from equilibrium, this relation no longer holds in its original form, motivating efforts to uncover generalized principles. One important direction has been the extension of the FDT to nonequilibrium systems~\cite{agarwal1972fluctuation,speck2006restoring,baiesi2009fluctuations,prost2009generalized,seifert2010fluctuation,altaner2016fluctuation}, where dissipation and dynamical activity play central roles in characterizing the response~\cite{baiesi2009fluctuations,seifert2010fluctuation}.
In parallel, thermodynamic and kinetic uncertainty relations (TUR and KUR) have revealed fundamental bounds on current fluctuations, highlighting intrinsic trade-offs between fluctuation and dissipation and between fluctuation and dynamical activity~\cite{barato2015thermodynamic,gingrich2016dissipation,garrahan2017simple,horowitz2020thermodynamic,van2022unified,Kwon_PRE2024, Lee_PRE2021}. These two lines of research on nonequilibrium fluctuations have advanced independently, without a clear link between them.

More recently, the response-TUR (R-TUR) was proposed as a conjecture for Markov jump processes~\cite{ptaszynski2024dissipation}, constraining the response of currents in terms of fluctuations and dissipation.
This conjecture was subsequently proven in the long-time limit through the discovery of fluctuation-response relations (FRRs)~\cite{aslyamov2025nonequilibrium}, identities that connect fluctuations with nonequilibrium responses to local perturbations of system parameters.
The R-TUR then follows as a corollary of these identities. Later, fluctuation-response inequalities (FRIs) were formulated, sharing the same structural form as the FRRs but appearing as finite-time inequalities that converge asymptotically to the FRRs~\cite{kwon2025fluctuation}.
Complementary studies, including the response kinetic uncertainty relation (R-KUR)~\cite{liu2025dynamical} and quantum generalizations~\cite{vu2025fundamental,kwon2025fluctuation}, have further broadened this framework. Yet, existing formulations of the FRRs and FRIs have been largely confined to Markov jump processes. Thus, although extending these relations to Langevin dynamics, the central paradigm for continuous-state nonequilibrium systems, has been repeatedly highlighted as an important direction~\cite{aslyamov2025nonequilibrium,ptaszynski2024nonequilibrium,ptaszynski2025nonequilibrium,liu2025dynamical}, it has remained largely unexplored. 

In this Letter, we establish a unified FRR for nonequilibrium Langevin dynamics in the long-time limit. This relation arises from a structural correspondence between fluctuations and responses to local perturbations of empirical density and current. By incorporating the local Onsager reciprocal relation that we derive in equilibrium, we demonstrate that the FRR reduces to the FDT in the equilibrium limit. Moreover, using the Cram\'er-Rao bound, we derive finite-time FRI that extends the connection between fluctuations and responses beyond the long-time limit and to arbitrary initial conditions. From the FRI, we further obtain response uncertainty relations, which provide experimentally accessible formulations based on global perturbations, as we shall demonstrate with the $F_1$-ATPase molecular motor model. These derivations complete the overall hierarchical relationship among the FRI, FRR, FDT, and TUR.

%%%%%%%%%%%%%%%%%%%%%%%%%%%%%%%%%%%%%%%%%%%%%%%%%%%%%%%%%%%%%%%%%%%%%%%%%%%%%
\emph{Setup}-- We formulate the main results in one-dimensional overdamped Langevin systems for clarity of presentation. This simple case already captures the essence of the FRR and FRI while avoiding technical complications. 
The extension to multidimensional settings is presented in Ref.~\cite{SM}. Thus, we consider a one-dimensional overdamped Langevin system described by
\begin{equation}\label{eq:Langevin}
    \dot{x}_t = \mu(x_t) F (x_t) + \sqrt{2 \mu(x_t) T(x_t)} \circledast \xi_t \ ,
\end{equation}
where $\circledast$ denotes the anti-It\^o product~\cite{lau2007state,anti-Ito}, $F(x)$ is an external force, and $\xi_t$ is a zero-mean Gaussian white noise with $\langle \xi_t \xi_s \rangle = \delta(t-s)$. $\mu(x)$ and $T(x)$ represent the position-dependent mobility and bath temperature, respectively. We set the Boltzmann constant to unity throughout.
The corresponding Fokker-Planck equation for the propagator $P(x,t|y,s)$ is given by
\begin{equation}\label{eq:FK_eq}
    \partial_t P(x,t|y,s) = \hat{\mathcal{L}}_x P(x,t|y,s),
\end{equation}
where $\hat{\mathcal{L}}_x \equiv -\partial_x \mu(x) [ F(x) - T(x) \partial_x ]$. Hereafter, the differential operators, such as $\hat{\mathcal{L}}_x$ and $\partial_x$, apply to all functions to their right, unless parentheses are used to restrict their range. For instance, $\partial_x f(x)g(x) = [\partial_x f(x)]g(x) + [\partial_x g(x)]f(x)$.
For notational convenience, we introduce the differential operator $\hat{\mathcal{J}}_x \equiv \mu(x) [ F(x) - T(x) \partial_x]$, so that $\hat{\mathcal{L}}_x = -\partial_x \hat{\mathcal{J}}_x$. We consider the following time-averaged observable: 
\begin{equation}
    \Theta(\tau) = \frac{1}{\tau}\int_0^\tau [f(x_t) + \dot{x}_t \circ g(x_t)] dt
\end{equation}
with arbitrary functions $f(x)$ and $g(x)$, where $\circ$ denotes the Stratonovich product~\cite{gardiner1985handbook}.
Its steady-state average is given by $\langle \Theta \rangle_{\rm ss} = \int [f(x) \pi(x) + g(x) j_{\rm ss}(x)] dx$, where $\pi(x)$ represents the steady-state probability distribution satisfying $\hat{\mathcal{L}}_x \pi(x) = 0$, and $j_{\rm ss}(x) \equiv \hat{\mathcal{J}}_x \pi(x)$ is the steady-state probability current~\footnote{Although the steady-state current $j_{\rm ss}(x)$ is spatially constant in the one-dimensional case, we keep the argument explicit to facilitate a clear connection to more general cases.}. 

\emph{Fluctuation-response relations}-- We consider first the response of $\langle \Theta \rangle_{\rm ss}$ to a local perturbation in a function $\phi(x)$ at $x=z$, expressed as the functional derivative $\delta \langle \Theta \rangle_{\rm ss}/\delta\phi(z)$. We restrict $\phi$ to the set $\{ \mu, F, T \}$, as these cases share a common structural form for the response function. A local perturbation in $\phi$ at position $z$ amounts to a shift of the differential operator $\hat{\mathcal{J}}_x$ into the form of $\hat{\mathcal{J}}_x + \varepsilon \delta(x-z) \hat{\mathcal{K}}_x^\phi$, where $\varepsilon$ is an infinitesimal factor and $\hat{\mathcal{K}}_x^\phi$ is an operator determined by the choice of $\phi$ (see Table~\ref{tab:K_and_N}). For two observables $\Theta_1$ and $\Theta_2$, our first main result--the FRR--relates the scaled covariance $C_{\Theta_1, \Theta_2} \equiv \lim_{\tau \to \infty} \tau {\rm Cov}\{ \Theta_1(\tau), \Theta_2(\tau) \}$ to the responses associated with two functions $\phi_1$ and $\phi_2$ through the identity
\begin{equation}\label{eq:FRR}
    C_{\Theta_1, \Theta_2}
    = \int \frac{2\pi(z) D(z)}{ N_{\phi_1}(z) N_{\phi_2}(z) }\frac{\delta \langle \Theta_1 \rangle_{\rm ss}}{\delta\phi_1(z)} \frac{\delta \langle \Theta_2 \rangle_{\rm ss}}{\delta\phi_2(z)} dz \; ,
\end{equation}
where $D(z) \equiv \mu(z) T(z)$ is the diffusion coefficient and $N_\phi(x) \equiv \hat{\mathcal{K}}_x^\phi \pi(x)$ (see Table~\ref{tab:K_and_N}).
This relation represents the continuum (Langevin) counterpart of the FRRs established for Markov jump processes~\cite{aslyamov2025nonequilibrium}. This single identity holds for all cross-combinations of $\phi_1$ and $\phi_2 \in\{ \mu, F, T\}$. It thus provides a generalization of the FDT, connecting global fluctuations of observables to their local responses in nonequilibrium Langevin systems. 

\begin{table}[t]
\renewcommand{\arraystretch}{1.3}
\centering
\begin{tabular}{|c||c|c|}
\hline
$\phi$ & $\hat{\mathcal K}_x^\phi$ & $N_\phi(x) = \hat{\mathcal K}_x^\phi \pi(x)$ \\
\hline\hline
$\mu$ & $F(x) - T(x)\,\partial_x$ & $j_{\rm ss}(x)/\mu(x)$ \\
$F$  & $\mu(x)$ & $\mu(x)\,\pi(x)$ \\
$T$  & $-\mu(x)\,\partial_x$ & $-\mu(x)\,\partial_x \pi(x)$ \\
\hline
\end{tabular}
\caption{Perturbation operators $\hat{\mathcal K}_x^\phi$ and prefactors $\mathcal N_\phi(x)$ for different choices of the perturbed function $\phi \in \{\mu, F, T\}$.}
\label{tab:K_and_N}
\end{table}

A brief outline of the derivation of Eq.~\eqref{eq:FRR} is given below.
Linear response analysis yields the following expressions (see Appendix A):
\begin{equation}\label{eq:response_ftns}
\begin{aligned}
    \frac{\delta \pi(x)}{\delta \phi(z)} = N_\phi(z) \partial_z H(x|z), \quad
    \frac{\delta j_{\rm ss}(x)}{\delta \phi(z)} = N_\phi(z) R(x,z)
\end{aligned}
\end{equation}
with $H(x|z) \equiv \int_0^\infty [P(x,t|z,0) - \pi(x)] dt$ and $R(x,z) \equiv \delta(x-z) + \hat{\mathcal{J}}_x\partial_z H(x|z)$.
Since the factor $N_\phi(z)$ is shared by both the response of $\pi(x)$ and that of $j_{\rm ss}(x)$, the ratio of the local response of a general observable $\Theta$ with respect to $N_\phi$ is independent of the choice of $\phi$, i.e., 
\begin{equation} \label{eq:response_to_N}
    \frac{1}{N_\phi(z)}\frac{\delta\langle \Theta \rangle_{\rm ss}}{\delta \phi(z)} =\int \big[f(x) \partial_z H(x|z) + g(x) R(x,z)\big] dx.
\end{equation}
To evaluate $C_{\Theta_1, \Theta_2}$, we reexpress $\Theta_i(\tau)$ in terms of the empirical density $\rho(x,\tau) \equiv \tau^{-1}\int_0^\tau \delta(x_t - x) dt$ and the empirical current $\jmath(x,\tau) \equiv \tau^{-1} \int_0^\tau \dot{x}_t \circ \delta(x_t-x) dt$ as
\begin{equation}
    \Theta_i(\tau) = \int [f_i(x)\rho(x,\tau) + g_i(x)\jmath(x,\tau)] dx.
\end{equation}
Then, $C_{\Theta_1, \Theta_2}$ can be written as
\begin{align} \label{eq:Covariance_expanded}
    &\iint dx dy \big[ f_1(x)f_2(y) C_{\rho(x), \rho(y)} + f_1(x)g_2(y) C_{\rho(x), \jmath(y)} \nonumber \\
    &  + g_1(x)f_2(y) C_{\jmath(x), \rho(y)} + g_1(x)g_2(y) C_{\jmath(x), \jmath(y)} \big]= C_{\Theta_1, \Theta_2}.
\end{align}
The scaled covariances $C_{\alpha(x), \beta(y)}$, where $\alpha, \beta \in \{ \rho, \jmath \}$, are evaluated as (see Appendix B for details)
\begin{equation} \label{eq:C_alpha_beta}
    C_{\alpha(x),\beta(y)} = \int 2\pi(z) D(z) A_{\alpha(x)} A_{\beta(y)} dz
\end{equation}
with $A_{\rho(x)} = \partial_z H(x|z)$ and $A_{\jmath(x)} = R(x,z)$. We can evaluate the right-hand side of Eq.~\eqref{eq:FRR} by substituting Eq.~\eqref{eq:response_to_N} into Eq.~\eqref{eq:FRR}. Likewise, the left-hand side can be obtained by inserting Eq.~\eqref{eq:C_alpha_beta} into Eq.~\eqref{eq:Covariance_expanded}. Comparing these two expressions confirms the FRR.

%%%%%%%%%%%%%%%%%%%%%%%%%%%%%%%%%%%%%%%%%%%%%%%%%%%%%%%%%%%%%%%%%%%%%%%%%%%%%
\emph{Reduction to the fluctuation-dissipation theorem}--
The FRR reduces to the FDT in the equilibrium limit. To demonstrate this, we take the observables to be the empirical currents, $\Theta_1(\tau) = \jmath(x,\tau)$ and $\Theta_2(\tau) = \jmath(y,\tau)$. In this case, the FRR~\eqref{eq:FRR} becomes $C_{\jmath(x), \jmath(y)} = \int 2\pi(z) D(z) R(x,z) R(y,z) dz$.
In equilibrium, we can show that the local Onsager reciprocal relation, $[\delta\langle \jmath(x) \rangle/\delta F(z)]_{\rm eq} = [\delta \langle \jmath(z)\rangle/\delta F(x)]_{\rm eq}$, holds (see Appendix C). This relation implies the symmetry $\pi_{\rm eq}(z) D(z) R_{\rm eq}(x,z)/T(z) = \pi_{\rm eq}(x) D(x)R_{\rm eq}(z,x) /T(x)$ which follows from Eq.~\eqref{eq:response_ftns} together with the identity $\langle \jmath(x) \rangle_{\rm ss} = j_{\rm ss}(x)$. Using this symmetry, along with the projection property $\int R(x,z) R(z,y) dz = R(x,y)$ (see Appendix A), the FRR for $C_{\jmath(x), \jmath(y)}$ in the equilibrium limit reduces, for spatially homogeneous temperature, to the standard linear relation between local fluctuation and response:
\begin{equation}\label{eq:local_FDT}
    C_{\jmath(x), \jmath(y)}^{\rm eq} = 2 T \left[ \frac{\delta \langle \jmath (x) \rangle}{\delta F(y)}\right]_{\rm eq}. 
\end{equation}
This equality is the FDT for local currents.
Multiplying Eq.~\eqref{eq:local_FDT} by $g_1(x)$ and $g_2(y)$ and integrating with respect to $x$ and $y$, we obtain equilibrium FDT for two currents $J_i = \int g_i(x) \jmath (x,\tau)dx$ $(i=1,2)$. For example, for a free Brownian particle, choosing $g_1(x) = g_2(y) = 1$ yields the Einstein relation $D_\infty = \mu_\infty T$, where $D_\infty = \lim_{\tau\to \infty}\langle [x(\tau)-x(0)]^2\rangle/2\tau$ is the long-time diffusion coefficient and $\mu_\infty = [\partial \langle \dot{x} \rangle/\partial F]_{\rm eq} $ is the linear-response mobility. Thus, the FRR can be regarded as the nonequilibrium generalization of the equilibrium FDT.

%%%%%%%%%%%%%%%%%%%%%%%%%%%%%%%%%%%%%%%%%%%%%%%%%%%%%%%%%%%%%%%%%%%%%%%%%%%%%
\emph{Fluctuation-response inequality}--
While the FRR offers an exact connection between fluctuation and response, it applies only in the $\tau \to \infty$ limit and under steady-state initial conditions. Analogous to the case of Markov jump processes~\cite{kwon2025fluctuation}, a complementary FRI can be formulated for Langevin dynamics, which is valid for finite observation times and arbitrary initial conditions.

To formulate the FRI, we consider a spatiotemporally localized perturbation of $\phi\in \{\mu ,F,T\}$, which amounts to shifting $\phi(x)$ to $\phi(x) + \varepsilon \delta (x-z)\delta(t-s)$ with an infinitesimal $\varepsilon$. The corresponding response is defined via the functional derivative
$\delta \langle \Theta(\tau) \rangle / \delta \phi(z,s)$. In the following, we first consider the case $\phi = F$, and then extend the result to the cases $\phi \in \{\mu , T\}$. 
The lower bound on the variance of an observable $\Theta(\tau)$ can be evaluated from a functional version of the Cram\'er-Rao bound
%~\footnote{This is a limiting case of the Cram\'er-Rao bound with infinitely many perturbation parameters, continuously distributed over space and time. Section II of \cite{SM} includes an alternative derivation of the FRI without invoking the Cram\'er-Rao bound.}, 
treating the force field $F(x,t)$ as the perturbation parameter (See Appendix D):
\begin{equation}\label{eq:CR-bound}
\begin{aligned}
    {\rm Var} [\Theta(\tau) ]
    \geq \iint dx dz \iint_0^\tau dt ds ~ \mathcal{I}^{-1}_F(x,t;z,s) & \\
    \times \frac{\delta \langle \Theta(\tau) \rangle}{\delta F(x,t)}  \frac{\delta \langle \Theta(\tau) \rangle}{\delta F(z,s)} \; ,
\end{aligned}
\end{equation}
where $\mathcal{I}_F(x,t;z,s)$ denotes the Fisher information kernel (see \eqref{eq:Fisher_info_def} for the explicit expression) and $\mathcal{I}^{-1}_F(x,t;z,s)$ is its functional inverse satisfying $\int dy \int_0^{\tau} dr~ \mathcal{I}_F(x,t;y,r) \mathcal{I}^{-1}_F(y,r;z,s) = \delta(x-z) \delta (t-s) $. For such a localized perturbation, the kernel simplifies to (see Appendix E)
\begin{equation}\label{eq:Fisher_info}
    \mathcal{I}_F(x,t;z,s) = \frac{\mu(x)p(x,t)}{2T(x)} \delta(x-z)\delta(t-s) \ ,
\end{equation}
where $p(x,t) $ is the probability density at time $t$ evolved from an arbitrary initial distribution $p_0(x)$.
Thus, $\mathcal{I}_F^{-1}(x,t;z,s) = 2T(x)/[\mu(x)p(x,t)] \delta(x-z)\delta(t-s)$.
Upon substituting this result into Eq.~\eqref{eq:CR-bound}, we obtain the FRI for the force perturbation:
\begin{equation}\label{eq:FRI_F}
    {\rm Var} [ \Theta(\tau) ] \geq \int dz \int_0^\tau ds ~ \frac{2T(z)}{\mu(z)p(z,s)} \left( \frac{\delta\langle \Theta(\tau) \rangle}{\delta F(z,s)} \right)^2 \ .
\end{equation}
Similar to the case of time-independent perturbations, the combination $[\tilde{N}_\phi (z,s) ]^{-1} \delta \langle \Theta(\tau) \rangle /\delta \phi (z,s) $, where $\tilde{N}_\phi(z,s) \equiv \hat{\mathcal{K}}_z^\phi p(z,s)$, is independent of the choice of $\phi$ (see Sec. II of Ref.~\cite{SM}).
This leads to
\begin{equation}\label{eq:FRI_general}
    {\rm Var} [ \Theta(\tau) ] \geq \int dz \int_0^\tau ds ~ \frac{2p(z,s)D(z)}{[\tilde{N}_\phi(z,s)]^2} \left( \frac{\delta\langle \Theta(\tau) \rangle}{\delta \phi(z,s)} \right)^2 \ .
\end{equation}
This  unified FRI provides a universal lower bound on the variance of any trajectory-dependent observable, in terms of its linear response to local perturbations.

For the steady-state initial condition, $p(z,s)$ and $\tilde{N}_\phi(z,s)$ reduce to $\pi(z)$ and $N_\phi(z)$, respectively. 
By applying the inequality $\int_0^\tau [\delta\langle \Theta(\tau)\rangle/\delta\phi(z,s)]^2 ds \geq \tau^{-1}[\int_0^\tau \delta\langle \Theta(\tau)\rangle/\delta\phi(z,s) ds ]^2 = \tau^{-1}[\delta\langle \Theta(\tau)\rangle/\delta\phi(z)]^2$ to Eq.~\eqref{eq:FRI_general}, we obtain
\begin{equation}\label{eq:FRI_ss}
    \tau {\rm Var} [ \Theta(\tau) ] \geq \int \frac{2\pi(z)D(z)}{[N_\phi(z)]^2} \left( \frac{\delta\langle \Theta(\tau) \rangle}{\delta \phi(z)} \right)^2 dz \;. 
\end{equation}
This inequality has the same structural form as the FRR~\eqref{eq:FRR} with $\Theta = \Theta_1 = \Theta_2$ and $\phi = \phi_1 = \phi_2$, but appears as an inequality rather than an equality. This inequality holds for any finite-time $\tau$ and becomes saturated in the limit $\tau \to \infty$.

%%%%%%%%%%%%%%%%%%%%%%%%%%%%%%%%%%%%%%%%%%%%%%%%%%%%%%%%%%%%%%%%%%%%%%%%%%%%%
\emph{Response uncertainty relation}--
The FRI provides a fundamental lower bound but requires full knowledge of the spatiotemporal response function $\delta\langle \Theta(\tau)\rangle/\delta\phi(z,s)$, which is typically inaccessible to experimental or numerical settings.
To obtain a more practical bound, though generally looser, we apply the Cauchy--Schwarz inequality to Eq.~\eqref{eq:FRI_general}. 
For an arbitrary function $\psi(z,s)$, we have the inequality $\iint [X(z,s)Y(z,s)]^2 \geq [\iint Y(z,s)]^2/\iint [1/X(z,s)]^2$ with $X(z,s) \equiv \sqrt{2p(z,s) D(z)}/[\psi(z,s) \tilde{N}_\phi(z,s)]$ and $Y(z,s) \equiv \psi(z,s) \delta\langle \Theta(\tau) \rangle/\delta \phi(z,s)$, where $\iint = \int dz \int_0^\tau ds$. 
Recognizing that $\delta_{\phi} \langle \Theta(\tau) \rangle \equiv \iint Y(z,s)$ corresponds to the response to the global perturbation $\phi(x,t) \mapsto \phi(x,t) + \varepsilon \psi(x,t)$, we obtain the response uncertainty relation:
\begin{equation}\label{eq:R-UR}
\frac{ [ \delta_\phi \langle \Theta (\tau) \rangle ]^2 }{ \rm{Var} [ \Theta(\tau) ] } 
\leq \frac{\psi_{\rm max}^2}{2} \int dz \int_0^\tau ds  ~ \frac{[\tilde{N}_\phi (z,s)]^2}{p(z,s)D(z)} \ ,
\end{equation}
where $\psi_{\rm max} = \sup_{x,t} |\psi(x,t)|$ is the maximum amplitude of the perturbation function over the domain of integration.
This inequality depends only on the response to a global perturbation, rather than the full local response profile, making it substantially more feasible than the FRI for practical implementation.

When $\phi  = F$, Eq.~\eqref{eq:R-UR} becomes
\begin{equation}
    \frac{ [ \delta_{F} \langle \Theta(\tau) \rangle ]^2 }{ \rm{Var} [ \Theta(\tau) ] } 
    \leq \psi_{\rm max}^2 A(\tau)  \ ,
\end{equation}
where $A(\tau) \equiv \int dx \int _0 ^\tau dt ~ \mu(x)p(x,t)/2T(x)$. 
This inequality represents the continuum analog of the R-KUR previously derived for Markov jump processes~\cite{liu2025dynamical,kwon2025fluctuation}, where the ratio of response to fluctuation is bounded by the dynamical activity.
While the dynamical activity diverges in the Langevin limit due to its inverse scaling with the microscopic length scale, this length scale factor is canceled in the continuum formulation by expressing the response as a functional derivative.

\begin{figure}
    \includegraphics[width=0.95\columnwidth]{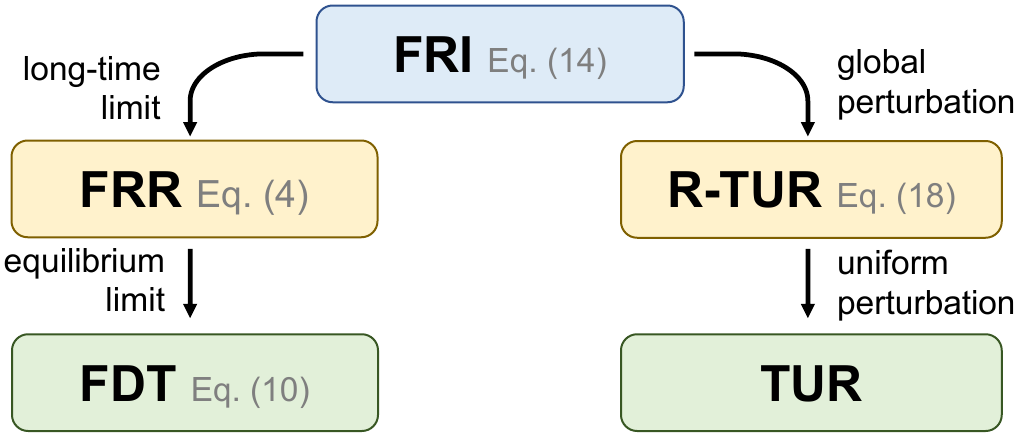}
    \caption{\label{fig:fig_connection}Hierarchy among the FRI, FRR, FDT, R-TUR, and TUR.} 
\end{figure}

A particularly meaningful choice of perturbation is $\phi = \ln\mu$. A perturbation in $\ln\mu$ represents a kinetic perturbation~\cite{chun2023trade,gao2024thermodynamic,gao2022thermodynamic}, which modifies the system's dynamics without altering thermodynamic forces in overdamped Langevin dynamics. This choice leads to $N_{\ln\mu}(x,t)=j(x,t)$, the probability current at time $t$, while preserving the theoretical framework developed so far.
Substituting this into Eq.~\eqref{eq:R-UR}, we obtain the R-TUR:
\begin{equation}
    \frac{ [ \delta_{\ln \mu} \langle \Theta(\tau) \rangle ]^2 }{ \rm{Var} [ \Theta(\tau) ] } 
    \leq \frac{\psi_{\rm max}^2 \Sigma_\tau}{2}  \ ,
\end{equation}
where $\Sigma_\tau = \int dx \int_0^\tau dt ~ [j(x,t)]^2/p(x,t)D(x)$ is the total entropy production~\cite{seifert2005entropy}.
The R-TUR establishes a thermodynamic trade-off between response, fluctuation, and dissipation.
Moreover, for current-like observables of the form $J(\tau) = \int g(x) \jmath(x,\tau) dx$, choosing a uniform perturbation $\psi(x,t) = \psi ~\forall(x, t)$ recovers the conventional TUR, $\langle (1+\tau\partial_\tau) J(\tau) \rangle^2/{\rm Var} [ J(\tau) ] \leq \Sigma_\tau/2$~\cite{koyuk2020thermodynamic} (see Sec.~III of Ref.~\cite{SM} for the derivation).
The hierarchical structure from the FRI to the R-TUR and finally to the TUR reveals that the FRI serves as a unifying response-theoretic foundation for existing uncertainty relations. It also establishes the theoretical connection between the FDT and the TUR in Langevin dynamics, as shown in Fig.~\ref{fig:fig_connection}.

%%%%%%%%%%%%%%%%%%%%%%%%%%%%%%%%%%%%%%%%%%%%%%%%%%%%%%%%%%%%%%%%%%%%%%%%%%%%%%
%%%%%%%%%%%%%%%%%%%%%%%%%%%%%%%%%%%%%%%%%%%%%%%%%%%%%%%%%%%%%%%%%%%%%%%%%%%%%%

%%%%%%%%%%%%%%%%%%%%%%%%%%%%%%%%%%%%%%%%%%%%%%%%%%%%%%%%%%%%%%%%%%%%%%%%%%%%%%
\begin{figure}
    \includegraphics[width=0.99\columnwidth]{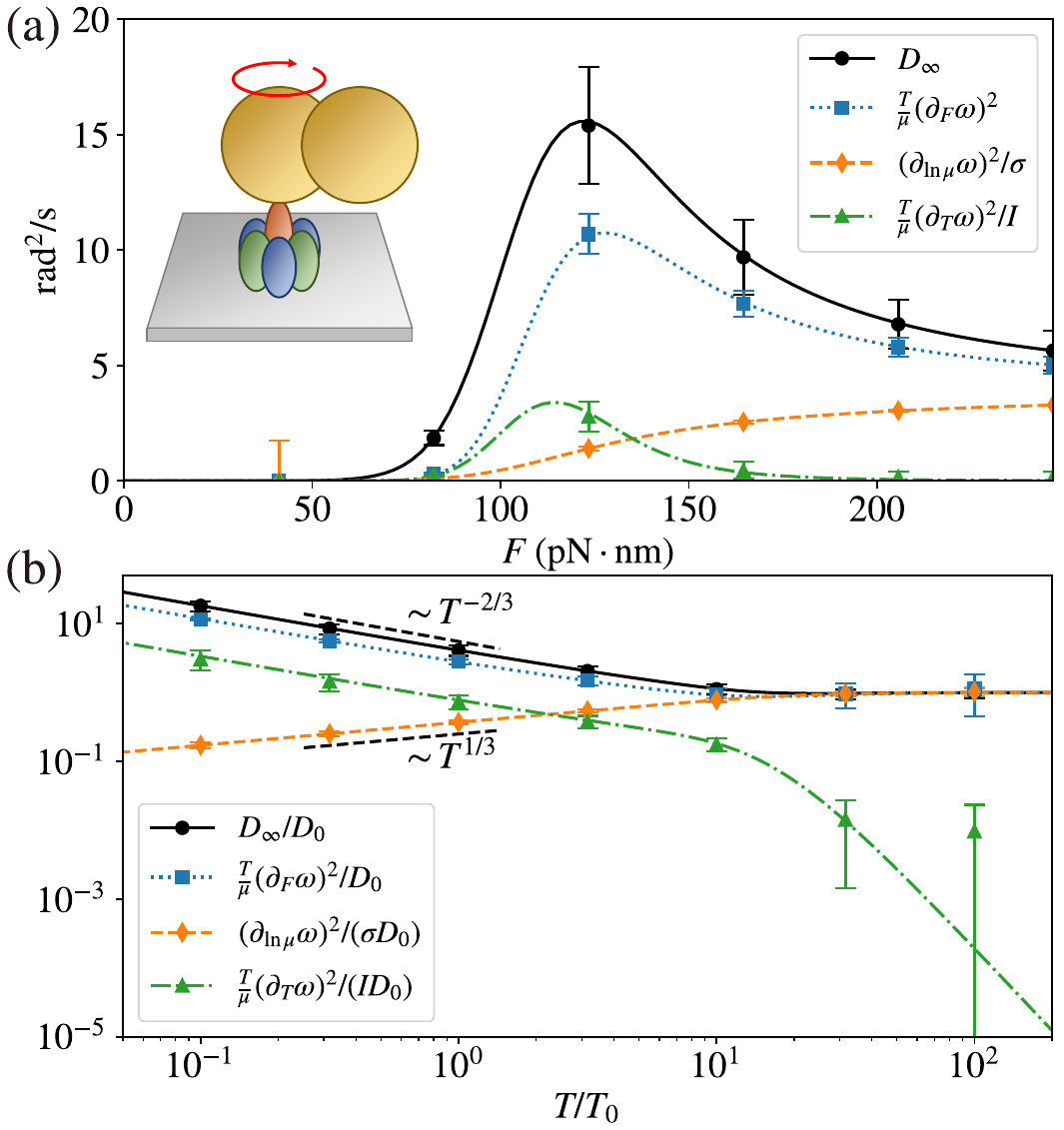}
    \caption{\label{fig:fig_F1-ATPase} Application of the response uncertainty relations to F$_1$-ATPase. (a) Comparison of $D_\infty$ with the three response-uncertainty bounds associated with perturbations in $F$, $\ln\mu$, and $T$.  
    Inset: Schematic of the experimental setup for F$_1$-ATPase attached to a duplex of polystyrene beads~\cite{Hayashi2015giant}.
    (b) Temperature dependence of $D_\infty /D_0$ at $F \approx 120\,\mathrm{pN\cdot nm}$, compared with the three bounds. $T_0=298\text{K}$ denotes the room temperature. In (a) and (b), solid, dotted, dashed, and dashdot lines represent the analytical results of $D_\infty$, $\frac{T}{\mu} (\partial_F \omega)^2$, $(\partial_{\ln \mu} \omega)^2 / \sigma$, and $\frac{T}{\mu}(\partial_T \omega)^2/I$, respectively, while markers with error bars (indicating the standard deviation) denote numerical simulation results obtained from Eq.~\eqref{eq:F1ATPase}.} 
\end{figure}
%%%%%%%%%%%%%%%%%%%%%%%%%%%%%%%%%%%%%%%%%%%%%%%%%%%%%%%%%%%%%%%%%%%%%%%%%%%%%%

\emph{Application to ${F}_1$-ATPase}--
We now apply the response uncertainty relations to a giant-diffusion phenomenon: the long-time diffusion coefficient of a Brownian particle moving in a tilted periodic potential can be enhanced by several orders of magnitude near the critical tilt of the potential~\cite{Reimann2001giant}. Since its discovery, this phenomenon has been extensively explored both theoretically~\cite{Reimann2002diffusion,Iida2025universality,Vo2025inverse} and experimentally~\cite{Lee2006giant,Hayashi2015giant,Kim2017giant}, particularly in studies of ${F}_1$-ATPase. 
The inset of Fig.~\ref{fig:fig_F1-ATPase}(a) illustrates the experimental setup used in Ref.~\cite{Hayashi2015giant}, where a duplex of polystyrene beads is attached to the $\gamma$ subunit of F$_1$-ATPase~\cite{ToyabePNAS2011,ToyabePRL2025}. An external electric field induces a constant dielectric torque on the rotor.
The dynamics of the cumulative angular displacement $\theta$, i.e., without applying modulo $2\pi$, is modeled by the overdamped Langevin equation~\cite{Hayashi2015giant},
\begin{equation}\label{eq:F1ATPase}
    \dot{\theta}_t= \mu [-\partial_\theta U(\theta)|_{\theta=\theta_t} + F] + \sqrt{2\mu T}\xi_t, 
\end{equation}
where $U(\theta)=U_0 \cos (3\theta)$ represents the periodic chemo-mechanical potential with $120^\circ$ rotational symmetry of the motor, $F$ is the externally applied torque, $\mu$ is the rotational mobility of the bead duplex, and $T$ is the ambient temperature.

Here, we consider steady-state initial conditions and a switch-on perturbation $\phi(x) \mapsto \phi(x) + \varepsilon h(t)$ where $h(t)$ is the Heaviside step function. The observable of interest is the time-averaged angular velocity, $\bar{\omega}(\tau) = \tau^{-1}\int_0^\tau \dot{\theta}_t dt$, whose variance determines the long-time diffusion coefficient through $D_\infty=\lim_{\tau\rightarrow\infty} {\rm Var}[\theta_\tau]/2\tau$. 
Applying Eq.~\eqref{eq:R-UR} to the cases $\phi = F$, $\ln\mu$, and $T$ in the limit $\tau\to \infty$, we obtain the following bounds on $D_\infty$,
\begin{equation}
\begin{aligned}
    D_\infty \geq \frac{T (\partial_F \omega )^2}{\mu} , \quad
    D_\infty \geq  \frac{(\partial_{\ln \mu } \omega )^2}{\sigma }, \quad
    D_\infty \geq \frac{T (\partial _T \omega)^2}{\mu I}
    \;,
\end{aligned}
\end{equation}
where $\omega \equiv \langle \bar{\omega} \rangle_{\rm ss}$ is the steady-state angular velocity, $\sigma \equiv \int_0^{2\pi} [j_{\rm ss}^2/\mu T\pi(\theta)] d\theta$ is the steady-state entropy production rate, and $I \equiv \int_0 ^{2\pi} [\pi '(\theta)]^2/\pi(\theta) d\theta = \langle [d_\theta S_{\rm sys} (\theta)]^2\rangle$.
Here $S_{\rm sys} (\theta) \equiv -\ln \pi(\theta)$ is the system entropy, and thus $I$ denotes the mean squared differential entropy. This setting is experimentally most relevant, as molecular-motor transport is typically probed under constant external driving and characterized via steady-state transport coefficients.

Figure~\ref{fig:fig_F1-ATPase}(a) compares $D_\infty$ with the three bounds.
The analytical results are supplemented by direct numerical simulations of Eq.~\eqref{eq:F1ATPase}, using parameter values ($\mu = 0.91 \text{rad/s/pN}\cdot\text{nm}$, $U_0 = 10k_\text{B}T$, $T = 298\text{K}$) reported in Ref.~\cite{Hayashi2015giant}. To better reflect experimental sampling limitations, 100 stochastic trajectories are simulated for each torque value, with each trajectory observed for a duration of $\tau=100$.
As $F$ increases, $D_\infty$ exhibits a pronounced peak near $F \approx 120\,\mathrm{pN\cdot nm}$, signaling the onset of the giant diffusion regime. 
Notably, the response functions with respect to $F$ and $T$ exhibit similar enhancement behaviors. To further assess the tightness of the bounds in the giant diffusion regime, we analyze their temperature dependence at the torque value corresponding to the diffusion peak, $F=120\,\mathrm{pN\cdot nm}$; see Fig.~\ref{fig:fig_F1-ATPase}(b).  
It is known that the ratio of the peak value of $D_\infty$ to the bare diffusion constant $D_0 = \mu T$ diverges as $T^{-2/3}$ in the low-temperature limit~\cite{Reimann2001giant}. 
The bounds associated with perturbations in $F$ and $T$ appear to follow this $T^{-2/3}$ scaling, tracking the divergence of the fluctuations. In contrast, the bound associated with $\mu$ scales with $T^{1/3}$, which vanishes at low temperatures, but becomes nearly as tight at higher temperatures as the torque-based bound.

\emph{Discussion}--
In this Letter, we established the FRR and the finite-time inequalities (the FRI, R-TUR, R-KUR, and TUR) for Langevin dynamics. The FRR provides a universal identity that connects fluctuations to linear response, both in and out of equilibrium, in the long-time limit under steady-state initial conditions.
The inequalities complement this identity by offering more practically accessible bounds that remain valid at finite observation times and for arbitrary initial conditions. From the FRI, we systematically recovered the R-KUR, the R-TUR, and finally the TUR, thereby unveiling a hierarchical structure underlying existing uncertainty relations. This hierarchy also elucidates the theoretical connection between the FDT and the TUR.

It is instructive to compare our work with two recent studies investigating fluctuation–response relations for Langevin dynamics~\cite{dechant2025finite,aslyamov2025macroscopic}. Both studies are formulated in the frequency domain, whereas our approach is entirely time-domain based. The FRI-type inequality in Ref.~\cite{dechant2025finite} saturates for linear systems irrespective of equilibrium, but does not reveal any explicit connection to the equilibrium FDT~\cite{dechant2025finite,aslyamov2025macroscopic}. In contrast, our framework naturally recovers the FDT in the equilibrium limit, regardless of the nonlinearity of the dynamics.
Furthermore, our FRI and response uncertainty relations apply not only to steady states but also to arbitrary initial conditions, and encompass a broad class of perturbations.

Extending our results to quantum stochastic processes, as done previously~\cite{Kwon_ComPhys2025}, represents a compelling direction for exploring the fundamental relationships among response, fluctuations, dissipation, and possibly quantum coherence in nonequilibrium systems.

\emph{Acknowledgements.}--
This research was supported by NRF Grants No.~2017R1D1A1B06035497 (H.P.), No.~RS-2023-00278985 (E.K.), and individual KIAS Grants No.~PG064902 (J.S.L.), PG089402 (H.-M.C.), and QP013602 (H.P.) at the Korea Institute for Advanced Study.
\\\\
H.-M.C. and E.K. contributed equally to this work.

\bibliography{main}

\section{End Matter}
\emph{Appendix A: Derivation of response functions}--The response functions to local perturbations can be derived using standard linear response theory~\cite{gardiner1985handbook,kubo2012statistical}.
We consider a local perturbation in $\phi \in \{ \mu, F, T\} $ at location $z$, implemented as $\hat{\mathcal{L}}_x \mapsto \hat{\mathcal{L}}_x - \varepsilon \partial_x \delta(x-z) \hat{\mathcal{K}}_x^\phi$ with an infinitesimal $\varepsilon$ and an operator $\hat{\mathcal{K}}_x^\phi$.
Assuming a steady-state initial condition, we write the probability distribution as $p(x,t) = \pi(x) + \varepsilon q(x,t;z) + \mathcal{O}(\varepsilon^2)$.
Inserting this expression into the Fokker-Planck equation $\partial_t p(x,t) = \hat{\mathcal{L}}_x p(x,t)$ and keeping terms of order $\varepsilon$ gives
\begin{equation}
    \frac{\partial q(x,t;z)}{\partial t}
    = \hat{\mathcal{L}}_x q(x,t;z) - \partial_x \delta(x-z)\hat{\mathcal{K}}_x^\phi \pi(x) \; ,
\end{equation}
and $q(x,0;z)=0$.
The propagator $P(x,t|y,s)$ is the solution of the unperturbed Fokker-Planck equation with initial condition $\lim_{t\to s}P(x,t|y,s) = \delta(x-y)$.
By the Green's function method, the solution of the linear differential equation for $q(x,t;z)$ is
\begin{equation}\label{eq:Q_expr1}
    q(x,t;z) = -\int_0^t ds \int dy ~ P(x,t|y,s) \partial_y \delta(y-z) \hat{\mathcal{K}}_y^\phi \pi(y) \;.
\end{equation}
Integrating by parts with respect to $y$ leads to
\begin{equation}\label{eq:Q_expr2}
\begin{aligned}
    q(x,t;z) = N_\phi(z) \int_0^t ds ~\partial_z P(x,t|z,s) \;.
\end{aligned}
\end{equation}
Using the fact that $\lim_{t\to \infty} q(x,t;z) = \delta \pi(x)/\delta \phi(z)$ and the definition of $H(x|z)$, we obtain the compact expression for the response function given in the first expression of Eq.~\eqref{eq:response_ftns}.
%Using the fact that $\lim_{t\to \infty} q(x,t;z) = \delta \pi(x)/\delta \phi(z)$ and the definition of $H(x|z)$ in Eq.~\eqref{eq:response_ftns}, we obtain a compact expression for the response function
%\begin{equation}
%    \frac{\delta \pi(x)}{\delta \phi(z)} = N_\phi(z) \partial_z H(x|z) \; .
%\end{equation}
The time-integrated propagator $H(x|y)$ has the following three properties, which are used throughout the derivations:
\begin{align}
    &\hat{\mathcal{L}}_y H(y|z) 
    = \pi(y) - \delta(y-z),
    \label{eq:property1_H} \\[4pt]
    &\hat{\mathcal{L}}^\dagger_z H(y|z) 
    = \pi(y) - \delta(y-z),
    \label{eq:property2_H} \\
    &\int H(y|z)\,\pi(z)\,dz 
    = 0.
\label{eq:property3_H}
\end{align}
The first two identities follow from time-integrating the forward and backward Fokker-Planck equations, $\partial_t P(y,t|z,0) = \hat{\mathcal{L}}_y P(y,t|z,0)$ and $\partial_t P(y,t|z,0) = \hat{\mathcal{L}}_z^\dagger P(y,t|z,0)$.
The last property is a consequence of stationarity, $\int P(y,t|z,0)\pi(z)dz = \pi(y)$.

Similarly, the probability current shifts as $j(x,t) = j_{\rm ss}(x) + \varepsilon k(x,t;z) + \mathcal{O}(\varepsilon^2)$, where $k(x,t;z) = \delta(x-z) \hat{\mathcal{K}}_x^\phi \pi(x) + \hat{\mathcal{J}}_x q(x,t;z)$. Taking the limit $t\to \infty$ and using the definition of $R(x,z)$ in Eq.~\eqref{eq:response_ftns} yields the second expression of Eq.~\eqref{eq:response_ftns}.
The function $R(x,y)$ satisfies $\int R(x,z) R(z,y) dz = R(x,y)$. This projection property holds in general and follows from the identity
\begin{align}\label{eq:RR_R}
    & \int dz ~ [\hat{\mathcal{J}}_x \partial_z H(x|z)] [\hat{\mathcal{J}}_z \partial_y H(z|y)] \nonumber \\
    & = \int dz ~ [\hat{\mathcal{J}}_x H(x|z)] [ \partial_y \hat{\mathcal{L}}_z H(z|y)] \\ 
    & = -\int dz ~ [\hat{\mathcal{J}}_x H(x|z)] [\partial_y \delta(z-y)] = - \hat{\mathcal{J}}_x \partial_y H(x|y), \nonumber
\end{align} 
where the first equality is obtained by integration by parts and the second equality uses Eq.~\eqref{eq:property1_H}.

%%%%%%%%%%%%%%%%%%%%%%%%%%%%%%%%%%%%%%%%%%%%%%%%%%%%%%%%%%%%%%%%%%%%%%%%%%
\emph{Appendix B: Derivations of Eq.~\eqref{eq:C_alpha_beta}}-- 
We begin by deriving the covariance of the empirical density, $\rho(x,\tau)=\tau^{-1}\int_0^\tau \delta(x_t - x) dt$.
Noting that ${\rm Cov} \{ \delta(x_t - x),  \delta(x_s - y) \}$ is given by $ P(x,t|y,s)\pi(y) - \pi(x) \pi(y)$ for $t>s$ and $P(y,s|x,t)\pi(x) - \pi(x) \pi(y)$ for $t<s$, we have
\begin{equation}
\begin{aligned}
    C_{\rho(x), \rho(y)}  &= \lim_{\tau \to \infty} \frac{1}{\tau} \int_0^\tau dt \int_0^t ds ~ 
    [P(x,t|y,s) - \pi(x) ] \pi(y) \\
    & + \lim_{\tau \to \infty} \frac{1}{\tau} \int_0^\tau dt \int_0^t ds ~ 
    [P(y,t|x,s) - \pi(y) ] \pi(x) \; ,
\end{aligned}
\end{equation}
where the integral variables are exchanged in the second integral.
Using the time-translation invariance $P(x,t|y,s) = P(x,t-s|y,0)$ and a change of integral variables $u\equiv t-s$, we can rewrite $\lim_{\tau\to\infty} \tau^{-1} \int_0^\tau dt \int_0^t ds$ as $\int_0^\infty du$ and thus obtain $C_{\rho(x), \rho(y)} = H(x|y)\pi(y) + H(y|x)\pi(x)$.

To connect the scaled covariance to the response functions, we rewrite $C_{\rho(x), \rho(y)}$ in terms of the local components $\partial_z H(x|z)$.
To this end, we use the two properties of $H$ in Eqs.~\eqref{eq:property2_H} and \eqref{eq:property3_H}, which allow us to obtain
\begin{equation}
\begin{aligned}
    H(x|y)\pi(y) & = \int  \pi(z) H(x|z) \delta(y-z) dz \\
    & = -\int \pi(z) H(x|z) \hat{\mathcal{L}}^\dagger_z H(y|z) dz \\
    & = -\int [\hat{\mathcal{J}}_z\pi(z) H(x|z)] \partial_z H(y|z) dz \\
    & = \int \pi(z) D(z) [ \partial_z H(x|z)] [\partial_z H(y|z)] dz \\
    & ~~~ - \int j_{\rm ss}(z) H(x|z) \partial_z H(y|z) dz \; ,
\end{aligned}
\end{equation}
where the first and second terms in the last expression are symmetric and antisymmetric under the exchange $x \leftrightarrow y$, respectively.
Adding the corresponding expression for $H(y|x)\pi(x)$ gives $H(x|y)\pi(y) + H(y|x) \pi(x) = \int 2\pi(z) D(z) [\partial_z H(x|z)][\partial_z H(y|z)] dz$.
Since $\partial_z H(x|z)$ can be written as $\partial_z H(x|z) = [N_\phi(z)]^{-1} \delta \pi(x)/\delta\phi(z)$ for any $\phi$, the empirical density satisfies the FRR
\begin{equation}\label{eq:C_rr}
    C_{\rho(x), \rho(y)}
    = \int \frac{2\pi(z) D(z)}{N_{\phi_1}(z) N_{\phi_2}(z)} \frac{\delta \pi(x)}{\delta \phi_1(z)} \frac{\delta \pi(y)}{\delta \phi_2(z)} dz
\end{equation}
valid for arbitrary choices of $\phi_1, \phi_2 \in \{ \mu, F, T\}$.

To evaluate the covariance of the empirical current, we first rewrite it as $\jmath(x,\tau) = \hat{\mathcal{J}}_x \rho(x,\tau) + \zeta(x,\tau)$, where the time-integrated noise $\zeta(x,\tau) = \tau^{-1} \int_0^\tau \sqrt{2D(x_t)}\xi_t \delta(x_t-x)$ satisfies $C_{\zeta(x,\tau), \zeta(y,\tau)}= 2 \pi(x) D(x) \delta(x-y)$ and $C_{\rho(x), \zeta(y)} = 2 \pi(y) D(y) \partial_y H(x|y)$ (see Sec.~I.B of \cite{SM} for details).
Using the identity $C_{\rho(x),\rho(y)} = \int 2\pi(z) D(z) [\partial_z H(x|z)][\partial_z H(y|z)] dz$ together with the definition $R(x,z) = \delta(x-z) + \hat{\mathcal{J}}_x \partial_z H(x|z)$, we obtain
\begin{align}
    &C_{\rho(x), \jmath(y)}
    = \int 2\pi(z)D(z) [\partial_z H(x|z)] R(y,z) dz \; , \label{eq:C_rj}\\
    &C_{\jmath(x), \jmath(y)} 
    = \int 2\pi(z)D(z) R(x,z) R(y,z) dz \; \label{eq:C_jj} .
\end{align}
Equation~\eqref{eq:C_alpha_beta} collects Eqs.~(\ref{eq:C_rr}--\ref{eq:C_jj}) into a single expression.
%Since $R(x,z) = [N_\phi(z)]^{-1} \delta j_{\rm ss}(x)/\delta\phi(z)$ for any $\phi \in \{\mu, F, T \}$, the scaled covariances involving empirical current also satisfy the FRR.
%This completes the proof that a general observable, being a linear combination of $\rho(x,\tau)$ and $\jmath(x,\tau)$, obeys the FRR in Eq.~\eqref{eq:FRR}.

%%%%%%%%%%%%%%%%%%%%%%%%%%%%%%%%%%%%%%%%%%%%%%%%%%%%%%%%%%%%%%%%%%%%%%%%%%
\emph{Appendix C: Local Onsager reciprocal relation}--
Here, we derive the local Onsager reciprocal relation $[\delta\langle \jmath(x) \rangle/\delta F(z)]_{\rm eq} = [\delta \langle \jmath(z)\rangle/\delta F(x)]_{\rm eq}$, which is equivalent to $\pi_{\rm eq}(z) \mu(z) R_{\rm eq}(x,z) = \pi_{\rm eq}(x) \mu(x)R_{\rm eq}(z,x)$, as follows from Eq.~\eqref{eq:response_ftns}.
Since the delta term in $R_{\rm eq}(x,z) = \delta(x-z) + \hat{\mathcal{J}}_x\partial_z H_{\rm eq}(x|z)$ trivially satisfies this symmetry, it suffices to show that
\begin{equation}\label{eq:Onsager_reciprocal1}
    \pi_{\rm eq}(z) \mu(z) \hat{\mathcal{J}}_x \partial_z H_{\rm eq}(x|z) = \pi_{\rm eq}(x) \mu(x) \hat{\mathcal{J}}_z \partial_x H_{\rm eq}(z|x) \; .
\end{equation}

In equilibrium, the absence of steady-state currents implies $ F(x)\pi_{\rm eq}(x) = T(x)\partial_x \pi_{\rm eq}(x)$, which yields $\hat{\mathcal{J}}_z H_{\rm eq}(x|z) \pi_{\rm eq}(z) = - \pi_{\rm eq}(z) D(z)\partial_z  H_{\rm eq}(x|z)$.
Applying $\hat{\mathcal{J}}_x$ to both sides gives 
\begin{equation}\label{eq:Onsager_reciprocal2}
    \hat{\mathcal{J}}_x \hat{\mathcal{J}}_z H_{\rm eq}(x|z) \pi_{\rm eq}(z)
= -\pi_{\rm eq}(z) D(z) \hat{\mathcal{J}}_x \partial_z  H_{\rm eq}(x|z) \;.
\end{equation}
By detailed balance, $H_{\rm eq}(x|z) \pi_{\rm eq}(z) = H_{\rm eq}(z|x)\pi_{\rm eq}(x)$, the left-hand side of Eq.~\eqref{eq:Onsager_reciprocal1} is invariant under exchanging $x \leftrightarrow z$.
Therefore, the right-hand side of Eq.~\eqref{eq:Onsager_reciprocal2} must also be invariant.
For a homogeneous temperature field $T(x)  = T(z) = T$, this reduces to Eq.~\eqref{eq:Onsager_reciprocal1}, which is equivalent to the local Onsager reciprocal relation.

%%%%%%%%%%%%%%%%%%%%%%%%%%%%%%%%%%%%%%%%%%%%%%%%%%%%%%%%%%%%%%%%%%%%%%%%%%
\emph{Appendix D: Functional version of the Cram\'{e}r-Rao bound}--
The Cram\'{e}r-Rao bound for multiple perturbation parameters $( \theta_1, \cdots, \theta_K )$ reads
\begin{equation} \label{eqA:CR_discrete}
    {\rm Var} [ \Theta(\tau) ] \geq \sum_{\alpha,\beta=1}^K \partial_{\theta_\alpha}\langle \Theta(\tau) \rangle [\mathcal{I}^{-1}(\tau)]_{\theta_\alpha \theta_\beta} \partial_{\theta_\beta}\langle \Theta(\tau) \rangle. 
\end{equation}
Here, $\mathcal{I}(\tau)$ denotes the Fisher information matrix, whose elements are given by $\mathcal{I}_{\theta_\alpha \theta_\beta}(\tau) = \langle \{\partial_{\theta_\alpha} \ln \mathcal{P}[\Gamma_\tau] \} \{\partial_{\theta_\beta} \ln \mathcal{P}[\Gamma_\tau] \} \rangle$~\cite{kwon2025fluctuation,kay1993fundamentals}, where $\mathcal{P}[\Gamma_\tau]$ is the probability of observing a stochastic trajectory $\Gamma_\tau$.
Equation~\eqref{eqA:CR_discrete} can be extended to the case where the perturbation parameter is a function defined on continuous space and time by discretizing the space-time domain into infinitesimal segments, such that the sums are replaced by integrals, i.e., $\sum_{\alpha} \to \int dx \int dt$. Accordingly, by replacing $\partial_{\theta_\alpha}$ with the functional derivative $\delta/\delta F(x,t)$, we immediately recover Eq.~\eqref{eq:CR-bound}.

%%%%%%%%%%%%%%%%%%%%%%%%%%%%%%%%%%%%%%%%%%%%%%%%%%%%%%%%%%%%%%%%%%%%%%%%%%
\emph{Appendix E: Derivation of the Fisher information kernel}--
Here we derive the explicit expression for the Fisher information kernel
\begin{equation}\label{eq:Fisher_info_def}
    \mathcal{I}_F (x,t;z,s) = \left\langle \frac{\delta \ln \mathcal{P}[\Gamma_\tau]}{\delta F(x,t)} \frac{\delta \ln \mathcal{P}[\Gamma_\tau]}{\delta F(z,s)} \right\rangle \;.
\end{equation}
To this end, we evaluate the path probability and compute the functional derivative of its logarithm with respect to the force field.
The probability of a trajectory $\Gamma_\tau = \{x_t\}_{t=0}^\tau$ is $\mathcal{P}[\Gamma_\tau] = \mathcal{N} p_0 (x_0) e^{-\mathcal{A}[\Gamma_\tau]}$, where $p_0(x)$ is the initial distribution, $\mathcal{A}[\Gamma_\tau]$ is the Onsager-Machlup action, and $\mathcal{N}$ is a normalization factor.
The action takes the form
\begin{equation}
    \mathcal{A}[\Gamma_\tau] = \int_0^\tau dt\, \frac{ [ \dot{x}_t - v (x_t) ]^2}{4 \mu (x_t) T (x_t)}
\end{equation}
where $v(x) = \mu(x) F(x) + [\partial_x \mu(x) T(x) ]$ is the effective drift including the spurious drift term arising from converting the anti-It\^o product to the It\^o product.
Taking the functional derivative of $\ln \mathcal{P}[\Gamma_\tau]$ with respect to $F(x,t)$ yields
\begin{equation}\label{eq:delta_lnP}
    \frac{\delta \ln \mathcal{P}[\Gamma_\tau]}{\delta F(x,t)} = \left\langle \frac{\delta \mathcal{A}[\Gamma_\tau]}{\delta F(x,t)} \right\rangle -\frac{\delta \mathcal{A}[\Gamma_\tau]}{\delta F(x,t)} \ ,
\end{equation}
where the identity $\delta \ln \mathcal{N} / \delta F(x,t) =  \langle \delta \mathcal{A}[\Gamma_\tau] / \delta F(x,t) \rangle$ is used.
The functional derivative of $\mathcal{A}[\Gamma_\tau]$ is given by
\begin{equation} \label{eq:deltaA}
    \frac{\delta \mathcal{A}[\Gamma_\tau]}{\delta F(x,t)} 
     = -\frac{\dot{x}_t - v(x_t) }{2 T(x_t)}  \delta (x - x_t) 
     = -\sqrt{\frac{\mu(x)}{2T(x)}} \xi_t \delta (x-x_t) ,
\end{equation}
where the second equality uses the Langevin equation \eqref{eq:Langevin}, which relates $\dot{x}_t - v (x_t)$ to the noise $\xi_t$. Thus, $\left\langle \delta \mathcal{A}[\Gamma_\tau]/\delta F(x,t) \right\rangle =0$.
Using Eqs.~\eqref{eq:delta_lnP} and \eqref{eq:deltaA}, we evaluate Eq.~\eqref{eq:Fisher_info_def} as
\begin{equation}
    \mathcal{I}_F(x,t;z,s) 
    =  \sqrt{ \frac{\mu(x)\mu(z)}{4 T(x) T(z)} } \left\langle \xi_t \xi_s 
    \delta(x - x_t) \delta(z - x_s) \right\rangle 
\end{equation}

Since the noise at the later time is independent of all variables determined by earlier noise realizations, the correlation
$\langle \xi_t \xi_s\, \delta(x-x_t)\delta(z-x_s)\rangle$ vanishes for $t\neq s$.
When $t = s$, the two noise values are delta-correlated, $\langle \xi_t \xi_s \rangle = \delta(t-s)$, and the noise at that time is independent of the state $x_t$.
Consequently, we obtain $\langle \xi_t \xi_s \delta ( x-x_t ) \delta ( z-x_s ) \rangle = \langle  \delta (x-x_t) \delta ( z-x_t ) \rangle \delta(t-s)= p(x,t) \delta(x-z)\delta(t-s)$, where $p(x,t) = \langle \delta (x-x_t) \rangle$ is the probability density at time $t$.
The Fisher information kernel therefore simplifies to Eq.~\eqref{eq:Fisher_info}.

\end{document}

\title{Supplemental Material --- Fluctuation-Response Theory for Nonequilibrium Langevin Dynamics}
\author{Hyun-Myung Chun}
\email{hmchun@kias.re.kr}
\affiliation{School of Physics, Korea Institute for Advanced Study, Seoul 02455, Republic of Korea}
\author{Euijoon Kwon}
\affiliation{School of Physics, Korea Institute for Advanced Study, Seoul 02455, Republic of Korea}
\affiliation{Department of Physics and Astronomy \& Center for Theoretical Physics, Seoul National University, Seoul 08826, Republic of Korea}
\author{Hyunggyu Park}
\affiliation{Quantum Universe Center, Korea Institute for Advanced Study, Seoul 02455, Korea}
\author{Jae Sung Lee}
\email{jslee@kias.re.kr}
\affiliation{School of Physics, Korea Institute for Advanced Study, Seoul 02455, Republic of Korea}

\maketitle

%%%%%%%%%%%%%%%%%%%%%%%%%%%%%%%%%%%%%%%%%%%%%%%%%%%%%%%%%%%%%%%%%%%%%%%%%%%%%
\section{Fluctuation-response relation for multidimensional overdamped Langevin systems}
In this section, we extend the derivation of the fluctuation-response relation (FRR) from the one-dimensional overdamped Langevin case presented in the main text to a general $N$-dimensional system.
Analogous to the one-dimensional case, we derive explicit expressions for the response functions and the scaled covariance matrix of the empirical density and empirical current.
Throughout this section, blackboard bold symbols ($\mathbb{A}$, $\mathbb{B}$, $\cdots$) are used to denote matrices (of dimension $N\times N$ or $N^2\times N$), while bold-faced italic symbols ($\bm{x}$, $\bm{y}$, $\cdots$) denote column vectors (of dimension $N \times 1$).
The transposes of bold-faced italic symbols ($\bm{x}^\textsf{T}$, $\bm{y}^\textsf{T}$, $\cdots$) denote the corresponding row vectors (of dimension $1 \times N$).
Furthermore, we define $\bm{\nabla}_{\bm{x}} \equiv  (\partial_{x_1}, \cdots, \partial_{x_N} )$ as a row-vector differential operator with respect to $\bm{x}$.

We consider an $N$-dimensional overdamped Langevin system described by
\begin{equation}\label{eq:Langevin}
\begin{aligned}
    \dot{\bm{x}}(t) 
    & = \mathbb{M}(\bm{x}(t))  \bm{F}(\bm{x}(t))
    + \sqrt{2} \mathbb{B}(\bm{x}(t)) \circledast\bm{\xi}(t) \\
    & = \bm{v}(\bm{x}(t)) + \sqrt{2} \mathbb{B}(\bm{x}(t)) \bullet \bm{\xi}(t),
\end{aligned}
\end{equation}
where $\bullet$ and $\circledast$ denotes the It\^o and anti-It\^o products, respectively, $\mathbb{M}(\bm{x})$ is the mobility matrix, $\bm{F}(\bm{x})$ is the force, $\mathbb{B}(\bm{x})$ is a square root of the diffusion matrix $\mathbb{D}(\bm{x}) = \mathbb{B}(\bm{x})  \mathbb{B}^{\textsf{T}}(\bm{x})$, and $\bm{\xi}(t)$ is a zero-mean Gaussian white noise with $\langle \xi_i(t) \xi_j(t') \rangle = \delta_{ij} \delta(t-t')$.
The second equality follows from converting the stochastic integration scheme from the anti-It\^o convention to the It\^o convention, yielding the effective drift
\begin{equation}
    \bm{v}(\bm{x}) = \mathbb{M}(\bm{x}) \bm{F}(\bm{x}) + \left[\bm{\nabla}_{\bm{x}} \mathbb{D}^\textsf{T} (\bm{x})\right] ^\textsf{T}.
\end{equation}
We assume that the mobility and diffusion matrices satisfy $\mathbb{D}(\bm{x}) = \mathbb{M}(\bm{x}) \mathbb{T}(\bm{x})$, where $\mathbb{T}(\bm{x}) = {\rm diag}\{ T_1, \cdots, T_N\}$ is a temperature matrix.
In the following, we assume that the diffusion matrix is symmetric, $\mathbb{D}^{\textsf{T}}(\bm{x}) = \mathbb{D}(\bm{x})$.
For later convenience, we also define a Gaussian noise
\begin{equation}
\bm{\eta}(t) = \sqrt{2}\, \mathbb{B}(\bm{x}(t)) \bullet \bm{\xi}(t),
\end{equation}
which satisfies $\langle \eta_i(t)\, \eta_j(t') \rangle = 2 D_{ij}(\bm{x}(t))\, \delta(t - t')$.

The corresponding Fokker–Planck equation for the propagator $P(\bm{x},t|\bm{y},s)$ reads
\begin{equation} \label{eq:Fokker_Planck}
    \partial_t P(\bm{x},t|\bm{y},s) = \hat{\mathcal{L}}_{\bm{x}} P(\bm{x},t|\bm{y},s)\;,
\end{equation}
with the differential operator
\begin{equation} \label{eq:Fokker_Planck2}
    \hat{\mathcal{L}}_{\bm{x}} = - \bm{\nabla}_{\bm{x}} \mathbb{M}(\bm{x})  [ \bm{F}(\bm{x}) - \mathbb{T}(\bm{x}) \bm{\nabla}_{\bm{x}} ^\textsf{T}] \;.
\end{equation}
We also introduce the associated probability current operator $\hat{\mathcal{J}}_{\bm{x}} = \mathbb{M}(\bm{x}) [ \bm{F}(\bm{x}) - \mathbb{T}(\bm{x}) \bm{\nabla}_{\bm{x}} ^\textsf{T} ]$.
We assume the existence of a unique steady-state distribution $\pi(\bm{x})$ satisfying $\hat{\mathcal{L}}_{\bm{x}}\pi(\bm{x}) = 0$, and denote the corresponding steady-state probability current as $\bm{j}^{\rm ss}(\bm{x}) = \hat{\mathcal{J}}_{\bm{x}} \pi(\bm{x})$.

To establish the FRR, we derive the linear responses of the empirical density and empirical current under local perturbations, as well as the scaled covariances of these observables. 
The empirical density is defined as $\rho(\bm{x},\tau) = \tau^{-1} \int_0^\tau \delta(\bm{x}(t)-\bm{x})\,dt$ and the empirical current as $\bm{\jmath}(\bm{x},\tau) = \tau^{-1} \int_0^\tau \dot{\bm{x}}(t) \circ \delta(\bm{x}(t)-\bm{x})\,dt$, whose steady-state averages are $\langle \rho(\bm{x},\tau)\rangle_{\rm ss} = \pi(\bm{x})$ and $\langle \bm{\jmath}(\bm{x},\tau) \rangle_{\rm ss} = \bm{j}^{\mathrm{ss}}(\bm{x})$, respectively.

\subsection{Linear response of the empirical density and current}
We consider a local perturbation in a system parameter $\phi \in \{ M_{ij}, F_i, T_i \}$ at position $\bm{z}$, implemented as $\hat{\mathcal{L}}_{\bm{x}} \mapsto \hat{\mathcal{L}}_{\bm{x}} - \varepsilon \bm{\nabla}_{\bm{x}}  \delta(\bm{x}-\bm{z}) \hat{\mathcal{K}}_{\bm{x}}^\phi$, where $\varepsilon$ is an infinitesimal parameter, and $\hat{\mathcal{K}}_{\bm x}^\phi$ is an operator determined by $\phi$.
The explicit forms of $\hat{\mathcal{K}}_{\bm{x}}^\phi$ for each type of perturbation are given by
\begin{equation}
    \hat{\mathcal{K}}_{\bm{x}}^{M_{ij}}
    = \mathbb{E}_{ij} [ \bm{F}(\bm{x}) - \mathbb{T}(\bm{x})  \bm{\nabla}_{\bm{x}}^\textsf{T}], \quad
    \hat{\mathcal{K}}_{\bm{x}}^{F_i}
    = \mathbb{M}(\bm{x}) \bm{e}_i, \quad
    \hat{\mathcal{K}}_{\bm{x}}^{T_i}
    = - \mathbb{M}(\bm{x})  \mathbb{E}_{ii} \bm{\nabla}_{\bm{x}}^\textsf{T},
\end{equation}
where $\mathbb{E}_{ij}$ is the matrix with zeros everywhere except for the $(i,j)$ component, which is 1, and $\bm{e}_i$ is the unit vector with 1 in the $i$-th component and zeros elsewhere.

Assuming steady-state initial condition, we write the probability distribution as $p({\bm x},t) = \pi(\bm{x}) + \varepsilon q(\bm{x},t;\bm{z}) + \mathcal{O}(\varepsilon^2)$.
Inserting into the Fokker-Planck equation $\partial_t p({\bm{x}},t) = \hat{\mathcal{L}}_{\bm{x}} p({\bm{x}},t)$ and keeping terms of order $\varepsilon$ gives
\begin{equation}
\frac{\partial q(\bm{x},t;\bm{z})}{\partial t}
= \hat{\mathcal{L}}_{\bm{x}} q(\bm{x},t;\bm{z}) - \bm{\nabla}_{\bm{x}}  \delta(\bm{x}-\bm{z})\hat{\mathcal{K}}_{\bm{x}}^\phi\pi(\bm{x}) \; ,
\end{equation}
and $q(\bm{x},0;\bm{z})=0$.
The propagator $P(\bm{x},t|\bm{y},s)$ is the solution of the unperturbed Fokker-Planck equation with the initial condition $\lim_{t\to s}P(\bm{x},t|\bm{y},s) = \delta(\bm{x}-\bm{y})$.
By the Green’s function method, the solution for $q(\bm{x},t;\bm{z})$ can be written as
\begin{equation}\label{eq:Q_expr1}
    q(\bm{x},t;\bm{z}) = -\int_0^t ds \int d\bm{y} ~ P(\bm{x},t|\bm{y},s) \bm{\nabla}_{\bm{y}}  \delta(\bm{y}-\bm{z}) \hat{\mathcal{K}}_{\bm{y}}^\phi \pi(\bm{y}) \;.
\end{equation}
Integrating by parts with respect to $\bm{y}$ leads to
\begin{equation}\label{eq:Q_expr2}
    q(\bm{x},t;\bm{z}) = \bm{N}_\phi ^\textsf{T}(\bm{z})  \int_0^t ds ~ \bm{\nabla}_{\bm{z}}^\textsf{T} P(\bm{x},t|\bm{z},s) \;,
\end{equation}
with $\bm{N}_\phi(\bm{z}) = \hat{\mathcal{K}}_{\bm{z}}^\phi \pi(\bm{z})$.
The explicit forms of the $m$th component of $\bm{N}_\phi(\bm{z})$ for each type of perturbation are given by
\begin{equation}
\begin{aligned}
    & [\bm{N}_{M_{ij}}(\bm{z})]_m
    = \delta_{im} [F_j(\bm{z}) - T_j(\bm{z}) \partial_{z_j}]\pi(\bm{z}), \\
    & [\bm{N}_{{F_i}}(\bm{z})]_m
    = M_{mi}(\bm{z}) \pi(\bm{z}), \quad
    [\bm{N}_{{T_i}}(\bm{z})]_m
    = - M_{mi}(\bm{z})\partial_{z_i} \pi(\bm{z}),
\end{aligned}
\end{equation}
Using the fact that $\lim_{t\to \infty} q(\bm{x},t;\bm{z}) = \delta \pi(\bm{x})/\delta \phi(\bm{z})$ and defining the convergent time-integrated propagator
\begin{equation}
    H(\bm{x}|\bm{z}) = \int_0^\infty [P(\bm{x},t|\bm{z},0) - \pi(\bm{x})] dt \; ,
\end{equation}
we obtain the compact expression for the response function of the probability density
\begin{equation}\label{eq:resp_density}
    \frac{\delta \pi(\bm{x})}{\delta \phi(\bm{z})} = \bm{N}_\phi^\textsf{T}(\bm{z}) \bm{\nabla}_{\bm{z}}^\textsf{T} H(\bm{x}|\bm{z}) \; .
\end{equation}
The time-integrated propagator has three properties:
\begin{equation}\label{eq:property1_H}
    \hat{\mathcal{L}}_{\bm{y}} H(\bm{y}|\bm{z}) = \pi(\bm{y}) - \delta(\bm{y}-\bm{z}),
\end{equation}
\begin{equation}\label{eq:property2_H}
    \hat{\mathcal{L}}^\dagger_{\bm z} H(\bm{y}|\bm{z}) = \pi(\bm{y}) - \delta(\bm{y}-\bm{z}),
\end{equation}
\begin{equation}\label{eq:property3_H}
    \int H(\bm{y}|\bm{z})\pi(\bm{z}) d\bm{z} = 0.
\end{equation}
The first two identities follow from time-integrating the forward and backward Fokker-Planck equations, $\partial_t P(\bm{y},t|\bm{z},0) = \hat{\mathcal{L}}_{\bm{y}} P(\bm{y},t|\bm{z},0)$ and $\partial_t P(\bm{y},t|\bm{z},0) = \hat{\mathcal{L}}_{\bm{z}}^\dagger P(\bm{y},t|\bm{z},0)$.
The last property is a consequence of stationarity, $\int P(\bm{y},t|\bm{z},0)\pi(\bm{z})d\bm{z} = \pi(\bm{y})$.

Similarly, the probability current can be expanded as $\bm{j}(\bm{x},t) = \bm{j}^{\rm ss}(\bm{x}) + \varepsilon \bm{k}(\bm{x},t;\bm{z}) + \mathcal{O}(\varepsilon^2)$ where $\bm{k}(\bm{x},t;\bm{z}) = \delta(\bm{x}-\bm{z}) \hat{\mathcal{K}}_{\bm{x}}^\phi \pi(\bm{x}) + \hat{\mathcal{J}}_{\bm{x}} q(\bm{x},t;\bm{z})$.
Taking $t\to \infty$ limit yields
\begin{equation}\label{eq:resp_current}
    \frac{\delta \bm{j}^{\rm ss}(\bm{x})}{\delta \phi(\bm{z})} = \mathbb{R}(\bm{x},\bm{z})  \bm{N}_\phi(\bm{z})\; 
\end{equation}
with $\mathbb{R}(\bm{x},\bm{z}) = \mathbb{I}\delta(\bm{x}-\bm{z}) + \hat{\mathcal{J}}_{\bm{x}} \bm{\nabla}_{\bm{z}}  H(\bm{x}|\bm{z})$.
The matrix $\mathbb{R}(\bm{x},\bm{y})$ satisfies the projection property $\int \mathbb{R}(\bm{x},\bm{z}) \mathbb{R}(\bm{z},\bm{y}) d\bm{z} = \mathbb{R}(\bm{x},\bm{y})$.
Denoting $\mathbb{I}(\bm{x},\bm{z}) \equiv \mathbb{I}\delta(\bm{x}-\bm{z})$ and $\mathbb{J}(\bm{x},\bm{z}) \equiv \hat{\mathcal{J}}_{\bm{x}} \bm{\nabla}_{\bm{z}} H(\bm{x}|\bm{z})$, the projection property follows from the identities $\int \mathbb{I}(\bm{x},\bm{z}) \mathbb{I}(\bm{z},\bm{y}) d\bm{z} = \mathbb{I}(\bm{x},\bm{y})$, $\int \mathbb{I}(\bm{x},\bm{z}) \mathbb{J}(\bm{z},\bm{y}) d\bm{z} = \int \mathbb{J}(\bm{x},\bm{z}) \mathbb{I}(\bm{z},\bm{y}) d\bm{z} = \mathbb{J}(\bm{x},\bm{y})$, and
\begin{equation}
\begin{aligned}
    \int d\bm{z} ~ [\mathbb{J}(\bm{x},\bm{z})]_{ij} [\mathbb{J}(\bm{z},\bm{y})]_{jk} 
    & = \sum_j \int d\bm{z} ~ [\hat{\mathcal{J}}_{\bm{x}} \partial_{z_j}  H(\bm{x}|\bm{z})]_i [\hat{\mathcal{J}}_{\bm{z}}  \partial_{y_k} H(\bm{z}|\bm{y})]_j \\
    & = \int d\bm{z} ~ [\hat{\mathcal{J}}_{\bm{x}} H(\bm{x}|\bm{z})]_i [\bm{\nabla}_{\bm{y}}^\textsf{T} \hat{\mathcal{L}}_{\bm{z}} H(\bm{z}|\bm{y})]_k \\
    & = -\int d\bm{z} ~ [\hat{\mathcal{J}}_{\bm{x}} H(\bm{x}|\bm{z})]_i [\bm{\nabla}_{\bm{y}} ^\textsf{T}\delta(\bm{z}-\bm{y})]_k \\ 
    & = - [\hat{\mathcal{J}}_{\bm{x}} \partial_{y_k} H({\bm x}|{\bm y})]_i
    = -[\mathbb{J}(\bm{x},\bm{y})]_{ik},
\end{aligned}
\end{equation}
where the second equality uses integration by parts and the third equality uses the property of $H$ in Eq.~\eqref{eq:property1_H}.

%%%%%%%%%%%%%%%%%%%%%%%%%%%%%%%%%%%%%%%%%%%%%%%%%%%%%%%%%%%%%%%%%%%%%%%%%%
\subsection{Scaled covariances of the empirical density and current}
We first derive the explicit expressions of the scaled covariances of empirical density, which is a time-integrated Dirac delta function.
Noting that ${\rm Cov} \{ \delta(\bm{x}(t) - \bm{x}),  \delta(\bm{x}(s) - \bm{y}) \}$ is given by $ P(\bm{x},t|\bm{y},s)\pi(\bm{y}) - \pi(\bm{x}) \pi(\bm{y})$ for $t>s$ and $P(\bm{y},s|\bm{x},t)\pi(\bm{x}) - \pi(\bm{x}) \pi(\bm{y})$ for $t<s$, we have
\begin{equation}
\begin{aligned}
    C_{\rho(\bm{x}), \rho(\bm{y})} 
    & = \lim_{\tau \to \infty} \frac{1}{\tau} \int_0^\tau dt \int_0^t ds ~ 
    [P(\bm{x},t|\bm{y},s) - \pi(\bm{x}) ] \pi(\bm{y}) \\
    & ~~~+ \lim_{\tau \to \infty} \frac{1}{\tau} \int_0^\tau dt \int_0^t ds ~ 
    [P(\bm{y},t|\bm{x},s) - \pi(\bm{y}) ] \pi(\bm{x}) \; ,
\end{aligned}
\end{equation}
where the integral variables are exchanged in the second integral.
Using the time-translation invariance $P(\bm{x},t|\bm{y},s) = P(\bm{x},t-s|\bm{y},0)$ and a change of integral variables $u\equiv t-s$, we can rewrite $\lim_{\tau\to\infty} \tau^{-1} \int_0^\tau dt \int_0^t ds$ as $\int_0^\infty du$ and thus obtain
\begin{equation}\label{eq:identity_rho}
   C_{\rho(\bm{x}), \rho(\bm{y})} = H(\bm{x}|\bm{y})\pi(\bm{y}) + H(\bm{y}|\bm{x})\pi(\bm{x}) \; .
\end{equation}

To connect the scaled covariance to the response functions, we rewrite $C_{\rho(\bm{x}), \rho(\bm{y})}$ in terms of the local components $\bm{\nabla}_{\bm{z}} H(\bm{x}|\bm{z})$.
To this end, we use the two properties of $H$ in Eqs.~\eqref{eq:property2_H} and \eqref{eq:property3_H}, from which we obtain
\begin{equation}
\begin{aligned}
    H(\bm{x}|\bm{y})\pi(\bm{y}) & = \int  \pi(\bm{z}) H(\bm{x}|\bm{z}) \delta(\bm{y}-\bm{z}) d\bm{z}
    = -\int \pi(\bm{z}) H(\bm{x}|\bm{z}) \hat{\mathcal{L}}^\dagger_{\bm{z}} H(\bm{y}|\bm{z}) d\bm{z} \\
    & = -\int [\hat{\mathcal{J}}_{\bm{z}}^\textsf{T} \pi(\bm{z}) H(\bm{x}|\bm{z})]  \bm{\nabla}_{\bm{z}}^\textsf{T} H(\bm{y}|\bm{z}) d\bm{z} \\
    & = \int \pi(\bm{z}) [\bm{\nabla}_{\bm{z}} H(\bm{x}|\bm{z})]  \mathbb{D}(\bm{z})  [\bm{\nabla}_{\bm{z}}^\textsf{T} H(\bm{y}|\bm{z})] d\bm{z} \\
    & \quad - \int [\bm{j}^{\rm ss}(\bm{z})]^{\textsf{T}} H(\bm{x}|\bm{z}) \bm{\nabla}_{\bm{z}}^\textsf{T} H(\bm{y}|\bm{z}) d\bm{z} \; ,
\end{aligned}
\end{equation}
where the first and second terms in the last expression are symmetric and antisymmetric under the exchange $\bm{x} \leftrightarrow \bm{y}$, respectively.
Adding the corresponding expression for $H(\bm{y}|\bm{x})\pi(\bm{x})$ yields 
\begin{equation}\label{eq:FRR_density_prototype}
    H(\bm{x}|\bm{y})\pi(\bm{y}) + H(\bm{y}|\bm{x}) \pi(\bm{x}) = \int 2\pi(\bm{z}) [\bm{\nabla}_{\bm{z}} H(\bm{x}|\bm{z})]  \mathbb{D}(\bm{z})  [\bm{\nabla}_{\bm{z}} ^\textsf{T}H(\bm{y}|\bm{z})] d\bm{z} \; .
\end{equation}

The term $\bm{\nabla}_{\bm{z}} H(\bm{x}|\bm{z})$ also appears in the expression of the response function of the empirical density in \eqref{eq:resp_density}.
To invert \eqref{eq:resp_density} and express $\bm{\nabla}_{\bm{z}} H(\bm{x}|\bm{z})$ in terms of the response functions, we explicitly label the spatial components of the perturbation parameters as $\phi_m(\bm{z})$, where $m \in \{ 1, \cdots ,N \}$ for $\phi = F$ and $T$, and $m \in \{ (11), (12), \cdots , (NN)\}$ for $\phi = M$.
Introducing a column vector $[\delta \pi(\bm{x})/\delta \bm{\phi}(\bm{z})]_m = \delta\pi(\bm{x})/\delta\phi_m(\bm{z})$, Eq.~\eqref{eq:resp_density} becomes
\begin{equation}
    \frac{\delta \pi(\bm{x})}{\delta \bm{\phi}(\bm{z})}
    = \mathbb{N}_{\bm{\phi}(\bm{z})}  \bm{\nabla}_{\bm{z}}^\textsf{T} H(\bm{x}|\bm{z}) \; ,
\end{equation}
where the matrix $\mathbb{N}_{\bm{\phi}}(\bm{z})$ is defined by $[\mathbb{N}_{\bm{\phi}}(\bm{z})]_{mn} = [\bm{N}_{\phi_m}(\bm{z})]_n$.
Assuming $\mathbb{N}_{\bm{\phi}}(\bm{z})$ is left-invertible, Eq.~\eqref{eq:FRR_density_prototype} leads to the FRR for the empirical density:
\begin{equation}
    C_{\rho(\bm{x}), \rho(\bm{y})}
    = \int 2\pi(\bm{z}) \left[\frac{\delta \pi(\bm{x})}{\delta \bm{\phi}^{(1)}(\bm{z})}\right]^\textsf{T}  [\mathbb{N}_{\bm{\phi}^{(1)}}^{-1}(\bm{z})]^\textsf{T} \mathbb{D}(\bm{z}) \mathbb{N}^{-1}_{\bm{\phi}^{(2)}}(\bm{z}) \frac{\delta \pi(\bm{y})}{\delta \bm{\phi}^{(2)}(\bm{z})} d\bm{z}
\end{equation}
valid for any combination of $\bm{\phi}^{(1)}, \bm{\phi}^{(2)} \in \{ \mathbb{M}, \bm{F}, \mathbb{T} \}$.
This generalizes the FRR for one-dimensional overdamped Langevin systems derived in the main text to multidimensional systems.
As an illustrative case, when $\bm{\phi}^{(1)} = \bm{\phi}^{(2)} = \bm{F}$ and thus $\mathbb{N}_{\bm{F}}^{-1}(\bm{z}) = (\mathbb{M}^\textsf{T})^{-1}(\bm{z})/\pi(\bm{z}) =  \mathbb{D}^{-1}(\bm{z}) \mathbb{T}(\bm{z})/\pi(\bm{z})$, the FRR reduces to
\begin{equation}
    C_{\rho(\bm{x}), \rho(\bm{y})}
    = \int \frac{2}{\pi(\bm{z})} \left[\frac{\delta \pi(\bm{x})}{\delta \bm{F}(\bm{z})}\right]^\textsf{T} \mathbb{T}(\bm{z}) \mathbb{D}^{-1}(\bm{z}) \mathbb{T}(\bm{z})  \frac{\delta \pi(\bm{y})}{\delta \bm{F}(\bm{z})} d\bm{z} \;,
\end{equation}
provided that $\mathbb{M}(\bm{z})$ is invertible.
The left-invertibility condition for the mobility perturbation is particularly stringent, since $\mathbb{N}_M(\bm{z})$ is a large rectangular ($N^2\times N$) matrix.
It is therefore useful to consider a reduced form of the FRR by noting that $\sum_j [\bm{N}_{\ln M_{ij}}(\bm{z})]_m = \sum_j M_{ij} [\bm{N}_{M_{ij}}(\bm{z})]_m = \delta_{im} j_i^{\rm ss}(\bm{z})$:
\begin{equation}
    C_{\rho(\bm{x}), \rho(\bm{y})}
    = \sum_{i,j,k,l} \int \frac{2\pi(\bm{z}) D_{ij}(\bm{z})}{j_i^{\rm ss}(\bm{z}) j_j^{\rm ss}(\bm{z})} \frac{\delta \pi(\bm{x})}{\delta \ln M_{ik}(\bm{z})} \frac{\delta \pi(\bm{y})}{\delta \ln M_{jl}(\bm{z})} d\bm{z} \; .
\end{equation}

Next, to evaluate the covariance of empirical current, we first rewrite it as
\begin{equation}
\begin{aligned}
    \bm{\jmath}(\bm{x},\tau)
    & = \frac{1}{\tau} \int_0^\tau \dot{\bm{x}}(t) \circ \delta(\bm{x} - \bm{x}(t)) dt \\
    & = \frac{1}{\tau} \int_0^\tau [\bm{v}(\bm{x}(t)) + \mathbb{D}(\bm{x}(t)) \bm{\nabla}_{\bm{x}(t)}^\textsf{T} + \bm{\eta}(t)  ] \bullet\delta(\bm{x} - \bm{x}(t)) dt \\
    & = \frac{1}{\tau} \int_0^\tau \left\{ \mathbb{M}(\bm{x}(t)) \bm{F}(\bm{x}(t))\delta(\bm{x}-\bm{x}(t))+ \left[\bm{\nabla}_{\bm{x}(t)}  \mathbb{D}(\bm{x}(t)) \delta(\bm{x}-\bm{x}(t))\right]^\textsf{T} \right\} dt \\
    & \quad + \underbrace{\frac{1}{\tau} \int_0^\tau \bm{\eta}(t) \bullet \delta(\bm{x} - \bm{x}(t)) dt}_{\equiv \bm{\zeta}(\bm{x},\tau)} \\
    & = \frac{1}{\tau} \int_0^\tau \left[ \mathbb{M}(\bm{x})\bm{F}(\bm{x}) - \mathbb{D}(\bm{x})\bm{\nabla}_{\bm{x}}^{\textsf{T}} \right] \delta(\bm{x} - \bm{x}(t))  dt + \bm{\zeta}(\bm{x},\tau) \\
    & = \hat{\mathcal{J}}_{\bm{x}} \rho(\bm{x},\tau) + \bm{\zeta}(\bm{x},\tau) \; .
\end{aligned}
\end{equation}
The time-integrated noise $\bm{\zeta}(\bm{x},\tau)$ satisfies
\begin{equation}
\begin{aligned}
    \langle \zeta_i(\bm{x},\tau) \zeta_j(\bm{y},\tau) \rangle_{\rm ss} 
    & = \frac{1}{\tau^2} \int_0^\tau dt \int_0^\tau dt'
    \left\langle 
    \eta_i(t) \eta_j(t') \delta(\bm{x}-\bm{x}(t)) \delta(\bm{y}-\bm{x}(t')) \right\rangle_{\rm ss} \\
    & = \frac{2}{\tau^2} \int_0^\tau dt
    ~ D_{ij}(\bm{x}) \delta(\bm{x}-\bm{y}) \langle \delta(\bm{x}-\bm{x}(t)) \rangle_{\rm ss} \\
    & = 2 \tau^{-1} \pi(\bm{x}) D_{ij}(\bm{x}) \delta(\bm{x}-\bm{y}) \;.
\end{aligned}
\end{equation}
We denote the scaled covariance matrix of two vector-valued observables $\bm{u}$ and $\bm{v}$ by $\mathbb{C}_{\bm{u}, \bm{v}}$, whose components are defined as $[\mathbb{C}_{\bm{u},\bm{v}}]_{ij} = C_{u_i,v_j}$.
With this notation, the scaled covariance matrix of the time-integrated noises is given by $\mathbb{C}_{\bm{\zeta}(\bm{x}), \bm{\zeta}^{\textsf{T}}(\bm{y})} = 2 \pi(\bm{x}) \mathbb{D}(\bm{x}) \delta(\bm{x} - \bm{y})$.
Similarly, we denote the scaled covariance of a scalar observable $a$ and a vector-valued observable $\bm{v}$ by $\bm{C}_{a,\bm{v}}$, which is a column vector with components $[\bm{C}_{a,\bm{v}}]_i = C_{a,v_i}$.
The scaled covariance between $\rho(\bm{x},\tau)$ and $\bm{\zeta}(\bm{y},\tau)$ is given by
\begin{equation}
\begin{aligned}
    \bm{C}_{\rho(\bm{x}), \bm{\zeta}(\bm{y})}& = \lim_{\tau \to \infty} \tau \langle \rho(\bm{x},\tau) \bm{\zeta}(\bm{y},\tau) \rangle_{\rm ss} \\
    & = \lim_{\tau \to \infty} \frac{1}{\tau} \int_0^\tau dt \int_0^\tau dt'
    ~ \langle \bm{\eta}(t') \delta(\bm{x}-\bm{x}(t)) \delta(\bm{y}-\bm{x}(t')) \rangle_{\rm ss} \\
    & = \lim_{\tau \to \infty} \frac{1}{\tau} \int_0^\tau dt \int_0^t dt'
    ~ \langle \bm{\eta}(t') \delta(\bm{x}-\bm{x}(t)) \delta(\bm{y}-\bm{x}(t')) \rangle_{\rm ss} \\
    & = \lim_{\tau \to \infty} \frac{2}{\tau} \int_0^\tau dt \int_0^t dt' ~\pi(\bm{y}) \mathbb{D}(\bm{y}) \bm{\nabla}_{\bm{y}}^\textsf{T} P(\bm{x},t|\bm{y},t') \\
    & = \lim_{\tau \to \infty}\frac{2}{\tau} \int_0^\tau du \int_0^{\tau-u} dt' ~\pi(\bm{y}) \mathbb{D}(\bm{y}) \bm{\nabla}_{\bm{y}}^\textsf{T} P(\bm{x},u|\bm{y},0) \\
    & = \lim_{\tau \to \infty} 2 \int_0^\tau du ~ \left( 1 - \frac{u}{\tau} \right) \pi(\bm{y}) \mathbb{D}(\bm{y}) \bm{\nabla}_{\bm{y}}^\textsf{T} P(\bm{x},u|\bm{y},0) \\
    & = 2 \int_0^\infty du ~ \pi(\bm{y}) \mathbb{D}(\bm{y}) \bm{\nabla}_{\bm{y}}^\textsf{T} P(\bm{x},u|\bm{y},0) \\
    & = 2 \pi(\bm{y}) \mathbb{D}(\bm{y}) \bm{\nabla}_{\bm{y}}^\textsf{T} H(\bm{x}|\bm{y})
\end{aligned}
\end{equation}
where the third equality uses causality, and the fourth equality uses
\begin{equation}
\begin{aligned}
    & \langle \bm{\eta}(t') \delta(\bm{x}-\bm{x}(t)) \delta(\bm{y}-\bm{x}(t')) \rangle_{\rm ss} \\
    & = \lim_{\Delta t \to 0} \frac{1}{\Delta t} \langle \bm{w}(t') \delta(\bm{x}-\bm{x}(t)) \delta(\bm{y}-\bm{x}(t')) \rangle_{\rm ss}  \\
    & = \lim_{\Delta t \to 0} \frac{1}{\Delta t}  \int d\bm{w} ~ \bm{w} P(\bm{x},t|\bm{y} + \bm{v}(\bm{y}) \Delta t + \bm{w} , t'+\Delta t) p_G(\bm{w}|\bm{y}) \pi(\bm{y})\\
    & = \lim_{\Delta t \to 0} \frac{1}{\Delta t}  \int d\bm{w} ~ \bm{w} p_G(\bm{w}|\bm{y}) \pi(\bm{y}) [(\bm{v}(\bm{y})\Delta t + \bm{w})^\textsf{T} \bm{\nabla}_{\bm{y}}^\textsf{T} + \Delta t \partial_{t'}]  P(\bm{x},t|\bm{y}, t')\\
    & = 2\pi(\bm{y}) \mathbb{D}(\bm{y}) \bm{\nabla}_{\bm{y}}^\textsf{T} P(\bm{x},t|\bm{y},t')
\end{aligned}
\end{equation}
with the Wiener increment $\bm{w}(t') = \int_{t'}^{t'+\Delta t} \bm{\eta}(s) ds$ and its conditional Gaussian distribution $p_G(\bm{w}|\bm{x})$.

Using the identities Eqs.~\eqref{eq:identity_rho} and \eqref{eq:FRR_density_prototype} together with the definition $\mathbb{R}(\bm{x},\bm{z}) = \mathbb{I} \delta(\bm{x}-\bm{z}) + \hat{\mathcal{J}}_{\bm{x}} \bm{\nabla}_{\bm{z}} H(\bm{x}|\bm{z})$, we obtain

\begin{equation}
\begin{aligned}
    & \bm{C}_{\rho(\bm{x}), \bm{\jmath}(\bm{y})}
    = \bm{C}_{\rho(\bm{x}), \hat{\mathcal{J}}_{\bm{y}} \rho(\bm{y})} + \bm{C}_{\rho(\bm{x}), \bm{\zeta}(\bm{y})}  \\
    & = \hat{\mathcal{J}}_{\bm{y}} \int  2\pi(\bm{z}) [\bm{\nabla}_{\bm{z}} H(\bm{y}|\bm{z})] \mathbb{D}(\bm{z})  [\bm{\nabla}_{\bm{z}}^\textsf{T} H(\bm{x}|\bm{z})] d\bm{z} + 2\pi(\bm{y}) \mathbb{D}(\bm{y}) \bm{\nabla}_{\bm{y}}^\textsf{T} H(\bm{x}|\bm{y}) \\
    & = \int 2\pi(\bm{z}) \mathbb{R}(\bm{y},\bm{z})  \mathbb{D}(\bm{z}) [\bm{\nabla}_{\bm{z}}^\textsf{T} H(\bm{x}|\bm{z})] d\bm{z} \; ,
\end{aligned}
\end{equation}
and
\begin{equation}
\begin{aligned}
    \mathbb{C}_{\bm{\jmath}(\bm{x}), \bm{\jmath}^\textsf{T}(\bm{y})} 
    & = \mathbb{C}_{\hat{\mathcal{J}}_{\bm{x}}\rho(\bm{x}), \hat{\mathcal{J}}_{\bm{y}}^\textsf{T} \rho(\bm{y})}  
    + \mathbb{C}_{\hat{\mathcal{J}}_{\bm{x}}\rho(\bm{x}), \bm{\zeta}^\textsf{T}(\bm{y})}  + \mathbb{C}_{\bm{\zeta}(\bm{x}), \hat{\mathcal{J}}_{\bm{y}}^\textsf{T} \rho(\bm{y})}  + \mathbb{C}_{\bm{\zeta}(\bm{x}), \bm{\zeta^\textsf{T}}(\bm{y})} \\
    & = \int 2\pi(\bm{z}) [\hat{\mathcal{J}}_{\bm{x}} \bm{\nabla}_{\bm{z}} H(\bm{x}|\bm{z})] \mathbb{D}(\bm{z}) [\hat{\mathcal{J}}_{\bm{y}} \bm{\nabla}_{\bm{z}} H(\bm{y}|\bm{z})]^\textsf{T} d\bm{z} \\
    & \quad + 2\pi(\bm{y}) [\hat{\mathcal{J}}_{\bm{x}} \bm{\nabla}_{\bm{y}} H(\bm{x}|\bm{y})] \mathbb{D}(\bm{y})  + 2\pi(\bm{x}) \mathbb{D}(\bm{x}) [\hat{\mathcal{J}}_{\bm{y}}  \bm{\nabla}_{\bm{x}} H(\bm{y}|\bm{x})]^\textsf{T} \\
    & \quad + 2\pi(\bm{x}) \mathbb{D}(\bm{x}) \delta(\bm{x}-\bm{y}) \\
    & = \int 2\pi(\bm{z}) \mathbb{R}(\bm{x},\bm{z}) \mathbb{D}(\bm{z})  \mathbb{R}^\textsf{T}(\bm{y},\bm{z}) d\bm{z} \; .
\end{aligned}
\end{equation}

Similar to the case of  the empirical density, the response function of the empirical current can also be expressed in terms of $\mathbb{R}$.
By inverting Eq.~\eqref{eq:resp_current} to write $\mathbb{R}$ in terms of the response function, we obtain that the scaled covariances involving empirical current also satisfy the same FRR structure.
This completes the proof that any observable that can be written as a linear combination of $\rho(\bm{x},\tau)$ and $\bm{\jmath}(\bm{x},\tau)$ obeys the FRR, just as in the one-dimensional case.

%%%%%%%%%%%%%%%%%%%%%%%%%%%%%%%%%%%%%%%%%%%%%%%%%%%%%%%%%%%%%%%%%%%%
\section{Derivation of the multidimensional fluctuation-response inequality}

In this section, we derive the multidimensional fluctuation-response inequality (FRI) that generalizes Eq.~(14) of the main text.
Unlike the derivation in the main text, we present here a derivation from first principles, without invoking the Cram\'er–Rao bound explicitly.
This provides a complementary perspective, while remaining mathematically equivalent to the approach used in the main text.

The path probability of observing a stochastic trajectory $\Gamma_\tau$ generated by the Langevin equation~\eqref{eq:Langevin} is given by $\mathcal{P}[\Gamma_\tau] = \mathcal{N} p_0(\bm{x}(0)) e^{-\mathcal{A}[\Gamma_\tau]}$, where $p_0(\bm{x})$ is the initial probability distribution, $\mathcal{N}$ is a normalization constant,  and
\begin{align}
        \mathcal{A}[\Gamma_\tau] = \frac{1}{4} \int_0 ^\tau dt ~ [\dot{\bm{x}}(t) - \bm{v}(\bm{x}(t))]^\textsf{T} \mathbb{D}^{-1}  [\dot{\bm{x}}(t) - \bm{v}(\bm{x}(t))]
    \;
\end{align}
is the Onsager-Machlup action.
Taking the functional derivative of $\ln \mathcal{P}[\Gamma_\tau]$ with respect to $F_i(\bm{z},s)$ yields
\begin{equation}
    \frac{\delta \ln \mathcal{P}[\Gamma_\tau]}{\delta F_i(\bm{z},s)} = \left\langle \frac{\delta \mathcal{A}[\Gamma_\tau]}{\delta F_i(\bm{z},s)} \right\rangle -\frac{\delta \mathcal{A}[\Gamma_\tau]}{\delta F_i(\bm{z},s)} \ ,
\end{equation}
where we used the identity $\delta \ln \mathcal{N} / \delta F_i(\bm{x},s) =  \langle \delta \mathcal{A}[\Gamma_\tau] / \delta F_i(\bm{x},s) \rangle$.
The functional derivative of $\mathcal{A}[\Gamma_\tau]$ is given by
\begin{equation}\label{eq:deltaA}
\begin{aligned}
    \frac{\delta \mathcal{A}[\Gamma_\tau]}{\delta F_i(\bm{z},s)} 
    & = - \frac{1}{2} \int_0^\tau dt ~[\dot{\bm{x}}(t) - \bm{v}(\bm{x}(t))]^\textsf{T} \mathbb{D}^{-1}(\bm{x}(t))  \frac{\delta \bm{v}(\bm{x}(t))}{\delta F_i(\bm{z},s)} \\
    & = -\frac{\dot{x}_i(s) -v_i(\bm{x}(s))}{2T_i(\bm{x}(s))} \delta(\bm{z} - \bm{x}(s)) \\
    & = - \sum_j \frac{B_{ij}(\bm{z})}{\sqrt{2}T_i(\bm{z})} \xi_j(s) \delta(\bm{z} - \bm{x}(s)) \ ,
\end{aligned}
\end{equation}
where the property $\mathbb{D}^{\textsf{T}}(\bm{x}) = \mathbb{D}(\bm{x}) = \mathbb{M}(\bm{x}) \mathbb{T}(\bm{x})$ has been used in the first equality and $\dot{\bm{x}}(t) - \bm{v}(\bm{x}(t))$ is related to the noise $\bm{\xi}(t)$ via the Langevin equation \eqref{eq:Langevin} in the second equality.
Since $\langle \xi_j (t) \rangle =0$, we obtain
\begin{align}
    Z_i (\bm{z},s) \equiv  \frac{\delta \ln \mathcal{P}[\Gamma_\tau]}{\delta F_i (\bm{z},s)} 
    = -\frac{\delta \mathcal{A}[\Gamma_\tau]}{\delta F_i(\bm{z},s)}
    = \sum_j \frac{B_{ij}(\bm{z})}{\sqrt{2}T_i(\bm{z})} \xi_j(s)\delta(\bm{z} - \bm{x}(s))
    \;.
\end{align}
The first moment of $Z_i(\bm{z},t)$ vanishes since $\langle \xi_j (t) \rangle =0$, while the second moment is given by
\begin{equation}
\begin{aligned}
    \langle Z_i (\bm{z},s) Z_j(\bm{z}', s') \rangle  
    & = \sum_{k,l} \frac{B_{ik}(\bm{z}) B_{jl}(\bm{z}')}{2T_i(\bm{z}) T_j(\bm{z}')}\left\langle \xi_k(s) \xi_l(s') \delta(\bm{z} - \bm{x}(s)) \delta(\bm{z}' - \bm{x}(s'))  \right\rangle \\
    & = \sum_{k,l} \frac{B_{ik}(\bm{z}) B_{jl}(\bm{z}')}{2T_i(\bm{z}) T_j(\bm{z}')} p(\bm{z},s) \delta_{kl} \delta(\bm{z}-\bm{z}') \delta(s-s')  \\
    & = \frac{D_{ij}(\bm{z})}{2T_i(\bm{z}) T_j(\bm{z})} p(\bm{z},s) \delta(\bm{z}-\bm{z}') \delta(s-s') \;,
\end{aligned}
\end{equation}
where $p(\bm{z},s) = \langle \delta(\bm{z} - \bm{x}(s)) \rangle$ is the probability density at time $s$ evolved from the initial distribution $p_0(\bm{x})$.

To derive the FRI, we use the fact that the following inequality holds for an arbitrary trajectory-dependent observable $\Theta(\tau)=\Theta[\Gamma_\tau]$ and a trajectory-independent vector ${\bm{G}}(\bm{z},s)$:
\begin{equation}
    \left\langle \left(\Theta(\tau) - \langle \Theta(\tau) \rangle - \int_0 ^\tau ds \int d\bm{z} ~ \bm{Z} ^\textsf{T}(\bm{z},s)  \bm{G}(\bm{z},s) \right)^2\right\rangle \ge 0 
    \;.
\end{equation}
Noting that
\begin{equation}
\begin{aligned}
    \left\langle \Theta(\tau) \bm{Z}^\textsf{T}(\bm{z},s) \bm{G}(\bm{z},s) \right\rangle
    & = \int \mathcal{D}\Gamma_\tau~ \mathcal{P}[\Gamma_\tau]\Theta[\Gamma_\tau] \left[\frac{\delta \ln \mathcal{P}[\Gamma_\tau]}{\delta \bm{F} (\bm{z},s)}\right]^\textsf{T}  \bm{G}(\bm{z},s) \\
    & = \left[ \frac{\delta}{\delta \bm{F} (\bm{z},s)} \int \mathcal{D}\Gamma_\tau~ \mathcal{P}[\Gamma_\tau] \Theta[\Gamma_\tau] \right]^\textsf{T}\bm{G}(\bm{z},s) \\
    & = \left[\frac{\delta \langle \Theta(\tau) \rangle}{\delta \bm{F}(\bm{z},s)}\right]^\textsf{T} \bm{G} (\bm{z},s)
    \;.
\end{aligned}
\end{equation}
and
\begin{equation}
\begin{aligned}
    & \left\langle \left( \int_0^\tau ds \int d\bm{z} 
    ~ \bm{Z}^\textsf{T}(\bm{z},s)\bm{G}(\bm{z},s) \right)^2 \right\rangle \\
    & = \int_0^\tau ds \int d\bm{z} \sum_{i,j=1}^N \frac{D_{ij} (\bm{z})}{2T_i(\bm{z}) T_j(\bm{z})} p(\bm{z},s) G_i(\bm{z},s)G_j(\bm{z},s) \\
    & = \frac{1}{2} \int_0^\tau ds \int d\bm{z} ~ p(\bm{z},s) \bm{G}^\textsf{T}(\bm{z},s)  \mathbb{T}^{-1}(\bm{z})  \mathbb{D}(\bm{z})  \mathbb{T}^{-1}(\bm{z}) \bm{G}(\bm{z},s)
    \;,
\end{aligned}
\end{equation}
we have the bound on the variance of $\Theta(\tau)$ as
\begin{equation}
\begin{aligned}
    & {\rm Var}(\Theta(\tau)) \\
    & \geq 2 \int_0^\tau ds \int d\bm{z} ~ \left[ \frac{\delta\langle \Theta(\tau)\rangle}{\delta \bm{F}(\bm{z},s)} 
    -\frac{1}{4}  p(\bm{z},s) \mathbb{T}^{-1}(\bm{z}) \mathbb{D}(\bm{z})  \mathbb{T}^{-1}(\bm{z})  \bm{G}(\bm{z},s)
    \right] ^\textsf{T}
    \bm{G}(\bm{z},s) \; .
\end{aligned}
\end{equation}
By choosing the vector $\bm{G}(\bm{z},s)$ as
\begin{equation}
    \bm{G}(\bm{z},s) = \frac{2}{p(\bm{z},s)} \mathbb{T}(\bm{z}) \mathbb{D}^{-1}(\bm{z}) \mathbb{T}(\bm{z}) \frac{\delta \langle \Theta(\tau) \rangle }{\delta \bm{F} (\bm{z},s)}
    \;,
\end{equation}
we arrive at the multidimensional FRI:
\begin{equation}\label{eq:FRI_for_f}
    \text{Var}(\Theta(\tau)) \ge 2 \int _0^\tau ds \int d\bm{z} ~ \frac{1}{p(\bm{z},s)}\left[\frac{\delta \langle \Theta(\tau) \rangle }{\delta \bm{F} (\bm{z},s)}\right]^\textsf{T} \mathbb{T}(\bm{z}) \mathbb{D}^{-1}(\bm{z}) \mathbb{T}(\bm{z}) \frac{\delta \langle \Theta(\tau) \rangle }{\delta \bm{F} (\bm{z},s)} \; .
\end{equation}

To connect the response to force perturbations with those to other types of perturbations that include a time-local aspect, i.e., $\phi(\bm{x}) \mapsto \phi(\bm{x}) +\varepsilon \delta(\bm{x}-\bm{z}) \delta(t-s)$, we reiterate the same linear-response analysis as in the previous section.
One readily finds
\begin{equation}
    \frac{\delta\langle \rho(\bm{x},\tau) \rangle}{\delta \phi(\bm{z},s)}
    = \frac{1}{\tau} \int_0^\tau  \frac{\delta p(\bm{x},t)}{\delta \phi(\bm{z},s)} dt 
    = \tilde{\bm{N}}_\phi(\bm{z},s)\bm{\nabla}_{\bm{z}}^\textsf{T} \tilde{H}_s(\bm{x}|\bm{z}) \;,
\end{equation}
and
\begin{equation}\label{eq:response_current}
    \frac{\delta \langle \bm{\jmath}(\bm{x},\tau) \rangle}{\delta \phi(\bm{z},s)}
    = \frac{1}{\tau} \int_0^\tau \frac{\delta j(\bm{x},t)}{\delta \phi(\bm{z},s)} dt = \tilde{\mathbb{R}}_s(\bm{x},\bm{z}) \tilde{\bm{N}}_\phi(\bm{z},s)\; 
\end{equation}
where $\bm{N}_\phi(\bm{z})$, $H(\bm{x}|\bm{z})$, and $\mathbb{R}(\bm{x},\bm{z})$ of Eqs.~\eqref{eq:resp_density} and \eqref{eq:resp_current} are replaced with $\tilde{\bm{N}}_\phi(\bm{z},s) = \hat{\mathcal{K}}_{\bm{z}}^\phi p(\bm{z},s)$,
\begin{equation}
    \tilde{H}_s(\bm{x}|\bm{z}) = \frac{1}{\tau} \int_s^{\tau} P(\bm{x},t|\bm{z},s) dt \;,
\end{equation}
and $\tilde{\mathbb{R}}_s(\bm{x},\bm{z}) = \tau^{-1} h(\tau - s) \mathbb{I} \delta(\bm{x}-\bm{z}) + \hat{\mathcal{J}}_{\bm{x}} \bm{\nabla}_{\bm{z}} \tilde{H}_s(\bm{x}|\bm{z})$, respectively.
The function $h(t)$ is the Heaviside step function.
Therefore, the combination $[\tilde{\mathbb{N}}_{\bm{\phi}} (\bm{z},s) ]^{-1} \delta \langle \Theta(\tau) \rangle /\delta \bm{\phi} (z,s)$ is independent of the choice of $\bm{\phi} \in \{ \mathbb{M}, \bm{F} ,\mathbb{T} \}$ as in the time-independent case for a steady-state initial condition.
Since $\tilde{\mathbb{N}}_{\bm{F}}(\bm{z},s) = \mathbb{M}^\textsf{T}(\bm{z})p(\bm{z},s) = \mathbb{T}^{-1}(\bm{z}) \mathbb{D}(\bm{z}) p(\bm{z},s)$,
\begin{equation}
    [\tilde{\mathbb{N}}_{\bm{\phi}} (\bm{z},s) ]^{-1} \frac{\delta \langle \Theta(\tau) \rangle}{\delta \bm{\phi}(\bm{z},s)}
    = \frac{1}{p(\bm{z},s)} \mathbb{D}^{-1}(\bm{z}) \mathbb{T}(\bm{z}) \frac{\delta\langle \Theta(\tau) \rangle}{\delta \bm{F}(\bm{z},s)} \;.
\end{equation}
for $\bm{\phi} \in \{ \mathbb{M}, \bm{F}, \mathbb{T}\}$.
Plugging this identity into \eqref{eq:FRI_for_f}, we have
\begin{equation}\label{eq:FRI_general}
    \text{Var}(\Theta(\tau)) \ge 2 \int _0^\tau ds \int d\bm{z} ~ p(\bm{z},s) \left[\frac{\delta \langle \Theta(\tau) \rangle }{\delta \bm{\phi} (\bm{z},s)}\right]^\textsf{T} [\tilde{\mathbb{N}}_{\bm{\phi}}^{\textsf{T}}(\bm{z},s)]^{-1} \mathbb{D}(\bm{z}) [\tilde{\mathbb{N}}_{\bm{\phi}}(\bm{z},s)]^{-1} \frac{\delta \langle \Theta(\tau) \rangle }{\delta \bm{\phi} (\bm{z},s)} \; ,
\end{equation}
which extends the applicability of the FRI in Eq.~(14) of the main text to multidimensional systems.

%%%%%%%%%%%%%%%%%%%%%%%%%%%%%%%%%%%%%%%%%%%%%%%%%%%%%%%%%%%%%%%%%%%%
\section{Derivation of the TUR from the R-TUR}
In this section, we show that the response thermodynamic uncertainty relation (R-TUR) [Eq.~(18) of the main text] can be used to directly derive the conventional thermodynamic uncertainty relation (TUR).

We first derive the response uncertainty relation for multidimensional overdamped Langevin systems.
For an arbitrary field $\bm{\psi}(\bm{z},s)$ with the same dimension as $\bm{\phi}(\bm{z},s)$ and a matrix $\mathbb{A}(\bm{z},s)$, the following Cauchy-Schwarz inequality holds:
\begin{equation}
\begin{aligned}
    \left[ \int_0^\tau ds \int d\bm{z} ~ \bm{\psi}^\textsf{T}(\bm{z},s) \frac{\delta \langle \Theta(\tau) \rangle}{\delta \bm{\phi}(\bm{z},s)} \right]^2
    & \leq \left[ \int_0^\tau ds \int d\bm{z} ~ \bm{\psi}^\textsf{T}(\bm{z},s) \mathbb{A}^{-1}(\bm{z},s) \bm{\psi}(\bm{z},s) \right] \\
    & \quad \times \left[ \int_0^\tau ds \int d\bm{z} ~ \left[\frac{\delta \langle \Theta(\tau) \rangle}{\delta \bm{\phi}(\bm{z},s)}\right]^\textsf{T} \mathbb{A}(\bm{z},s) \frac{\delta \langle \Theta(\tau) \rangle}{\delta \bm{\phi}(\bm{z},s)} \right] \;, 
\end{aligned}
\end{equation}
assuming that $\mathbb{A}(\bm{z},s)$ is symmetric positive definite.
The left-hand side corresponds to the response to the global perturbation $\bm{\phi}(\bm{x},t) \mapsto \bm{\phi}(\bm{x},t) + \varepsilon \bm{\psi}(\bm{x},t)$, which we denote by $\delta_{\phi}\langle \Theta(\tau) \rangle$.
Choosing $\mathbb{A}(\bm{z},s) = 2p(\bm{z},s) [\tilde{\mathbb{N}}_\phi^\textsf{T}(\bm{z},s)]^{-1} \mathbb{D}(\bm{z}) [\tilde{\mathbb{N}}_\phi(\bm{z},s)]^{-1}$, so that the second term on the right-hand side coincides with that of Eq.~\eqref{eq:FRI_general}, we obtain
\begin{equation}\label{eq:R-UR}
\begin{aligned}
    \frac{ [ \delta_{\phi} \langle \Theta (\tau) \rangle ]^2 }{ \rm{Var} [ \Theta(\tau) ] } 
    & \leq  \int_0^\tau ds \int d\bm{z} ~ \frac{1}{2p(\bm{z},s)}\bm{\psi}^\textsf{T}(\bm{z},s) \tilde{\mathbb{N}}_\phi(\bm{z},s) \mathbb{D}^{-1}(\bm{z}) \tilde{\mathbb{N}}_\phi^\textsf{T}(\bm{z},s) \bm{\psi}(\bm{z},s) \\
    & \leq \frac{\psi_{\rm max}^2}{2} \int_0^\tau ds \int d\bm{z} ~ \frac{1}{p(\bm{z},s)}\sum_{i,j}\left| [\tilde{\mathbb{N}}_\phi(\bm{z},s) \mathbb{D}^{-1}(\bm{z})  \tilde{\mathbb{N}}_\phi^\textsf{T}(\bm{z},s) ]_{ij} \right| \; ,
\end{aligned}
\end{equation}
where $\psi_{\rm max} \equiv \sup_{i,\bm{z},s} |\psi_i(\bm{z},s)|$.
In particular, when $\mathbb{M}(\bm{z}) = {\rm diag} \{ \mu_1, \cdots ,\mu_N \}$, the kinetic perturbation $\mu_i(\bm{x},t) \mapsto \mu_i(\bm{x},t) [ 1+ \varepsilon \psi_i(\bm{x},t)]$~\cite{gao2022thermodynamic,gao2024thermodynamic} leads to the R-TUR,
\begin{equation}\label{eq:R-TUR}
\begin{aligned}
    \frac{ [ \delta_{\ln 
    \mu} \langle \Theta (\tau) \rangle ]^2 }{ \rm{Var} [ \Theta(\tau) ] } 
    & \leq \frac{\psi_{\rm max}^2}{2} \int_0^\tau ds \int d\bm{z} ~ \sum_{m=1}^N\frac{ [j_m(\bm{z},s)]^2}{\mu_m(\bm{z}) T_m(\bm{z}) p(\bm{z},s)}
    = \frac{\psi_{\rm max}^2}{2} \Sigma_{\tau}\; ,
\end{aligned}
\end{equation}
where $\Sigma_\tau = \int_0^\tau ds \int d\bm{z} \sum_{m=1}^N [j_m(\bm{z},s)]^2/[\mu_m(\bm{z})T_m(\bm{z}) p(\bm{z},s)]$ is the total entropy produced up to time $\tau$~\cite{seifert2005entropy}.

For current-like observables of the form $J(\tau) = \int \bm{g}^\textsf{T}(\bm{x})  \bm{\jmath}(\bm{x},\tau) d\bm{x}$, choosing a uniform kinetic perturbation $\psi_i(\bm{x},t) = \psi ~\forall(i, \bm{x}, t)$ recovers the conventional TUR.
From Eq.~\eqref{eq:response_current}, the response to the uniform perturbation is computed as
\begin{equation}
\begin{aligned}
    \delta_{\ln \mu} \langle J (\tau) \rangle
    & = \psi \int_0^\tau  ds \int d\bm{z}  \sum_{m=1}^N \frac{\delta}{\delta \ln \mu_m(\bm{z},s)} \int d\bm{x} ~ \bm{g}^\textsf{T}(\bm{x}) \langle \bm{\jmath}(\bm{x},\tau) \rangle \\
    & = \psi \int_0^\tau  ds \int d\bm{x} \int d\bm{z} ~ \bm{g}^\textsf{T}(\bm{x}) \tilde{\mathbb{R}}_s(\bm{x},\bm{z})  \bm{j}(\bm{z},s) \\
    & =  \frac{\psi}{\tau}  \int_0^\tau ds \int d\bm{z} ~ \bm{g}^\textsf{T}(\bm{z})  \bm{j}(\bm{z},s) \\
    & \quad +  \frac{\psi}{\tau} \int_0^\tau ds \int_s^\tau dt \int d\bm{z} \int d\bm{x} ~ \bm{g}^\textsf{T}(\bm{x}) [\hat{\mathcal{J}}_{\bm{x}} \bm{\nabla}_{\bm{z}} P(\bm{x},t|\bm{z},s) ]  \bm{j}(\bm{z},s) \\
    & =  \psi \langle J(\tau) \rangle
    + \frac{\psi}{\tau} \int_0^\tau ds \int d\bm{x} ~ \bm{g}^\textsf{T}(\bm{x}) \hat{\mathcal{J}}_{\bm{x}} [ p(\bm{x},\tau) - p(\bm{x},s) ] \\
    & =  \psi \langle J(\tau) \rangle
    + \frac{\psi}{\tau} \int_0^\tau ds \int d\bm{x} ~ \bm{g}^\textsf{T}(\bm{x})  [ \bm{j}(\bm{x},\tau) - \bm{j}(\bm{x},s) ] \\
    & =  \psi ( 1 + \tau \partial_\tau)  \langle J(\tau) \rangle \;,
\end{aligned}
\end{equation}
where the following chain of identities has been used in the fourth equality:
\begin{equation}
\begin{aligned}
    & \int_0^\tau ds \int_s^\tau dt \int d\bm{z} ~ \bm{j}^\textsf{T}(\bm{z},s)\bm{\nabla}_{\bm{z}} ^\textsf{T}P(\bm{x},t|\bm{z},s) \\
    & = -\int_0^\tau ds \int_s^\tau dt\int d\bm{z} ~ P(\bm{x},t|\bm{z},s)  \bm{\nabla}_{\bm{z}}  \bm{j}(\bm{z},s) \\
    & = \int_0^\tau ds \int_s^\tau dt \int d\bm{z} ~ P(\bm{x},t|\bm{z},s) \partial_s p(\bm{z},s) \\
    & = -\int_0^\tau ds \int_s^\tau dt \int d\bm{z} ~ p(\bm{z},s) \partial_sP(\bm{x},t|\bm{z},s) \\
    & = \int_0^\tau ds \int_s^\tau dt \int d\bm{z} ~ p(\bm{z},s) \partial_t P(\bm{x},t|\bm{z},s) \\
    & = \int_0^\tau ds ~  [ p(\bm{x},\tau) - p(\bm{x},s) ] \;.
\end{aligned}
\end{equation}
Plugging $\delta_{\ln \mu}\langle J(\tau)\rangle = \psi(1 + \tau \partial_\tau) \langle J(\tau) \rangle$ and $\psi_{\rm max} = \psi$ into Eq.~\eqref{eq:R-TUR}, we thus recover the conventional TUR~\cite{koyuk2020thermodynamic}
\begin{equation}
    \frac{[(1 + \tau\partial_\tau)\langle J(\tau)\rangle]^2}{{\rm Var}[J(\tau)]} \leq \frac{\Sigma_\tau}{2} \;.
\end{equation}